\documentstyle[12pt,epsf]{article}

\begin{document}

\baselineskip=18pt plus 0.2pt minus 0.1pt
\parskip = 6pt plus 2pt minus 1pt

\makeatletter

\@addtoreset{equation}{section}
\renewcommand{\theequation}{\thesection.\arabic{equation}}

\renewcommand{\thefootnote}{\fnsymbol{footnote}}

\newcommand{\CR}[1]{\left[ #1 \right\}}
\newcommand{\bm}[1]{\mbox{\boldmath $#1$}}
\newcommand{\calD}{{\cal D}}
\newcommand{\abs}[1]{\left\vert #1\right\vert}
\newcommand{\p}{\partial}
\newcommand{\VEV}[1]{\left\langle #1\right\rangle}
\newcommand{\braket}[2]{\VEV{#1 | #2}}
\newcommand{\oxi}{\widehat\xi}
\newcommand{\oeta}{\widehat\eta}
\newcommand{\sgn}{\mathop{\rm sgn}}
\newcommand{\ds}{\displaystyle}
\renewcommand{\Re}{\mathop{\rm Re}}
\renewcommand{\Im}{\mathop{\rm Im}}
\newcommand{\Half}{\frac{1}{2}}
\newcommand{\dTlambda}{\delta_{\{T,\lambda\},\{T',-\lambda'\}}}
\newcommand{\bra}[1]{\left\langle #1 \right|}
\newcommand{\ket}[1]{\left| #1 \right\rangle}
\newcommand{\rbra}[1]{{}_{\lower.5ex\hbox{$\scriptstyle\rm R$}}
                     {\left\langle #1 \right|}}
\newcommand{\mbra}[1]{{}_{\lower.5ex\hbox{$\scriptstyle\rm M$}}
                     {\left\langle #1 \right|}}
\newcommand{\rket}[1]{{{\left| #1 \right\rangle}_{\rm R}}}
\newcommand{\mket}[1]{{{\left| #1 \right\rangle}_{\rm M}}}
\newcommand{\QB}{Q_{\rm B}}
\newcommand{\tQB}{\widetilde{Q}_{\rm B}}
\newcommand{\bQB}{\bm{Q}_{\rm B}}
\newcommand{\dB}{\delta_{\rm B}}
\newcommand{\calV}{{\cal V}}
\newcommand{\mt}{\mu_\bot}
\newcommand{\nn}{\nonumber}
\newcommand{\hanbi}[2]{\frac{\delta #1}{\delta #2}}
\newcommand{\cald}{{\cal D}}
\newcommand{\ac}{{\overline{c}}}
\newcommand{\henbi}[2]{\frac{\partial #1}{\partial #2}}
\newcommand{\til}[1]{{\widetilde{#1}}}
\newcommand{\sayuu}[1]{{\stackrel{\leftrightarrow}{#1}}}

\makeatother

\begin{titlepage}
  \title{ \hfill
\parbox{4cm}{\normalsize KUNS-1374\\HE(TH)~95/24\\hep-th/9512206}\\
\vspace{1cm}
String Field Theory in Rindler Space-Time and String Thermalization
}
\author{
Hiroyuki Hata\thanks{e-mail address: {\tt
  hata@gauge.scphys.kyoto-u.ac.jp}}
  {}\thanks{Supported in part by Grant-in-Aid for Scientific
  Research from Ministry of Education, Science and Culture
  (\#06640390).}, {}
Hajime Oda\thanks{e-mail address: \tt
  oda@gauge.scphys.kyoto-u.ac.jp} {} and
Shigeaki Yahikozawa\thanks{e-mail address: \tt
  yahiko@gauge.scphys.kyoto-u.ac.jp}
  {\,}\thanks{Supported in part by Grant-in-Aid for Scientific
  Research from Ministry of Education, Science and Culture
  (\#07854012).}
\\
{\normalsize\em Department of Physics, Kyoto University}\\
{\normalsize\em Kyoto 606-01, Japan}}
\date{\normalsize December, 1995}
\maketitle
\thispagestyle{empty}

\begin{abstract}
\normalsize

Quantization of free string field theory in the Rindler
space-time is studied by using the covariant formulation and
taking the center-of-mass value of the Rindler string time-coordinate
$\eta(\sigma)$ as the time variable for quantization.
We construct the string Rindler modes which vanish in either
of the Rindler wedges $\pm$ defined by the Minkowski center-of-mass
coordinate of the string.
We then evaluate the Bogoliubov coefficients between
the Rindler string creation/annihilation operators and
the Minkowski ones, and analyze the string thermalization.
An approach to the construction of the string Rindler modes
corresponding to different definitions of the wedges is also presented
toward a thorough understanding of the structure of the Hilbert space
of the string field theory on the Rindler space-time.

\end{abstract}
\end{titlepage}

\section{Introduction}
\label{sec:intro}

{}From various viewpoints including the problem of information loss and
thermalization in the black hole space-time, of much importance is
to study string theory near the event horizon of
a black hole \cite{tHooft,SusskindUglum}.
Instead of the real black hole, it is simpler and instructive to
consider the Rindler space-time; a flat space-time
$(ds)^2=-\xi^2(d\eta)^2+(d\xi)^2+(dx^\bot)^2$  with an event horizon
($x^\bot$ denotes the transverse space coordinates).
In fact, in the large mass limit of a black hole, the space-time
near the event horizon approaches the Rindler space-time.

The purpose of this paper is to study the string theory in the Rindler
space-time on the basis of string field theory (SFT), which is the most
natural and convenient framework for investigating the problems
concerning the string creations and annihilations.
In particular, we employ the covariant SFT based on the BRST
formulation \cite{Siegel,FreeSFT,WittenSFT,HIKKOopen}, which is more
suited to the present subject than the light-cone
SFT \cite{KakuKikkawa}.
Although the Rindler space-time is flat and the quantization of
local (particle) field theory on it is well-known \cite{BD,bible}, the
quantization of SFT on the Rindler space-time is totally a non-trivial
matter due to the fact that string is an extended object and the
Rindler space-time has a horizon.
(The Rindler quantization of string theory
using the first quantized formalism was previously considered
in \cite{VegaSanchez,LoweStrominger}.)
In this paper, we consider the free open bosonic SFT in 26 dimensions,
and ignore the problem related to the presence of the tachyon mode.

In the Rindler quantization of SFT, the string field is
a functional of the Rindler string coordinates
$\left(\xi(\sigma),\eta(\sigma),X^\bot(\sigma)\right)$ related to the
Minkowski string coordinates $X^\mu(\sigma)$ by
\begin{equation}
X^0(\sigma)=\xi(\sigma)\sinh\eta(\sigma),
\quad
X^1(\sigma)=\xi(\sigma)\cosh\eta(\sigma)\ .
\label{eq:eta(s)_xi(s)}
\end{equation}
The SFT action expressed in terms of
$\left(\xi(\sigma),\eta(\sigma),X^\bot(\sigma)\right)$
(see eq.\ (\ref{eq:S_R}) in Sec.\ 4.1) has an invariance under the
$\sigma$-independent shift of the $\eta(\sigma)$ coordinate, and we
shall take the center-of-mass (CM) coordinate of $\eta(\sigma)$,
\begin{equation}
\eta_0=\int_0^\pi\frac{d\sigma}{\pi}\eta(\sigma)\ ,
\label{eq:eta_0}
\end{equation}
as the Rindler time for quantization.

\begin{figure}[htbp]
\begin{center}
\leavevmode
\epsfxsize=8cm
\epsfbox{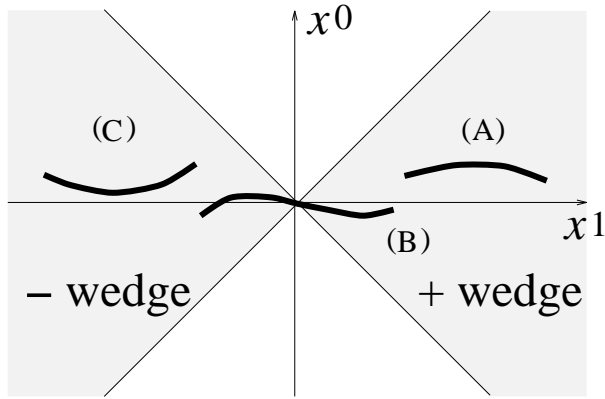}
\caption{The three strings (A), (B) and (C). They are confined in the
Rindler wedges (shaded region), and string (B) crosses the origin.}
\label{fig:wedge}
\end{center}
\end{figure}

The most important problem in the Rindler quantization of
SFT is the definition of Hilbert spaces of states on each Rindler
wedge.
The whole Rindler space-time can be divided into two Rindler wedges
$\pm$ depicted in Fig.\ 1, where the Rindler wedge $+$ ($-$) are
defined by $x^1 > |x^0|$  $(x^1 < -|x^0|)$.
In a local (particle) field theory, a point in one Rindler wedge is
not causally connected to a point in the other wedge, so the
whole Hilbert space on the Rindler space-time can be considered as a
product of two Hilbert spaces on the Rindler wedges $\pm$.
In the case of SFT, however, there exist configurations of a
string ((B) in Fig.\ \ref{fig:wedge}) which crosses the origin and
lives in both Rindler wedges, and this makes the definition of the
Hilbert spaces on the Rindler wedges a difficult and obscure problem.
In fact, it is not clear whether the string (B) in Fig.\
\ref{fig:wedge} is causally connected to strings (A) and (C)
in Fig.\ \ref{fig:wedge}.
(It is evident that the string (A) in the $+$ wedge is causally
disjoint to the string (C) in the $-$ wedge.)

To completely clarify the structure of the Hilbert space of SFT on the
Rindler space-time, we need deep understanding on the causality of
strings in the Rindler space-time. In spite of recent investigations
on the causality of strings \cite{Martinec,Love,LPSTU},
we do not yet have a complete solution.
In this paper, we first adopt one way of defining the Hilbert
spaces on each Rindler wedge.
This is to define the division of the whole Hilbert space into
two by whether the Minkowski CM coordinate of the string,
$x^\mu_{\rm CM}=\int_0^\pi\left(d\sigma/\pi\right)X^\mu(\sigma)$,
is in the Rindler wedge $+$ or $-$.
In Sections 3 and 4, we shall apply this definition to determine a
complete orthonormal system of the string Rindler modes on each
Rindler wedge and evaluate the Bogoliubov coefficients between the
Minkowski string creation/annihilation operators and the Rindler ones,
and study the string thermalization.

This division of the Hilbert space, however, is not a fully
satisfactory one in the point that the division is through the
Minkowski CM coordinate of the string.
Division by the Rindler CM coordinate,
$\xi_0=\int_0^\pi \left(d\sigma/\pi\right)\xi(\sigma)$,
would be more natural.
Although the complete analysis of the construction of the string
Rindler modes corresponding to various ways of dividing the Hilbert
space is still beyond our technical ability, in Sec.\ 5 we shall
present a construction of the string Rindler modes corresponding to
different divisions than the one adopted in Secs.\ 3 and 4.
We hope that this construction will shed light on the complete
understanding of the wedge problem.

Before closing the Introduction, let us mention a naive approach to the
Rindler quantization of SFT.
This is to first regard the SFT on Minkowski space-time as a set
of the infinite number of component (particle) field theories with
masses and spins lying on the Regge trajectories, and then to carry
out the Rindler quantization of the respective component fields.
Since a component field $\varphi(x^\mu_{\rm CM})$ is a function of the
Minkowski CM coordinate $x^\mu_{\rm CM}$ of the string, it is this CM
string coordinate $x^\mu_{\rm CM}$ rather than the whole string
coordinate $X^\mu(\sigma)$ that we put on the Rindler space-time in
this approach.
The Rindler time for this quantization is $\oeta$ defined by
\begin{equation}
x^0_{\rm CM}=\oxi\sinh\oeta,\qquad x^1_{\rm CM}=\oxi\cosh\oeta\ ,
\label{eq:oeta_oxi}
\end{equation}
and is different from $\eta_0$ of eq.\ (\ref{eq:eta_0}).
Note that a choice of the Rindler time for quantization leads to the
corresponding definition of the positive frequency modes of string
wave function and consequently it determines the structure of the
Rindler ground state. Therefore, the string thermalizations could
differ between the Rindler quantization using the $\eta_0$ time and
the one using the $\oeta$ time.

\begin{figure}[htbp]
\begin{center}
\leavevmode
\epsfxsize=8cm
\epsfbox{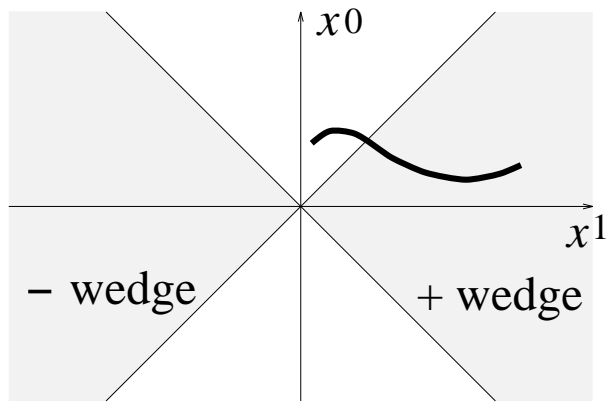}
\caption{A string with $x^\mu_{\rm CM}$ in the $+$ wedge but crossing
the Rindler horizon.}
\label{fig:hamidasi}
\end{center}
\end{figure}

However, this naive approach is unsatisfactory and unnatural
since the ``Rindler string'' here is not always confined in the
Rindler space-time.
For example, there exists a ``Rindler string'' depicted in Fig.\
\ref{fig:hamidasi}:
its CM coordinates $x^\mu_{\rm CM}$ is within Rindler
space-time but a part of the ``Rindler string'' is outside of it.
That is, this ``Rindler string'' crosses the Rindler horizon.
Although we do not adopt this approach in this paper, we shall
in some places make comparison with this another
``Rindler quantization'' method by referring it as ``naive approach''.

The organization of the rest of this paper is as follows.
First summarizing in Sec.\ 2 the quantization of free SFT on the
Minkowski space-time, in Sec.\ 3 we construct the string Rindler modes
corresponding to the above explained division of the Hilbert
space. Using this Rindler modes, we carry out the Rindler quantization
of free SFT and study the string thermalization in Sec.\ 4.
In Sec.\ 5, an attempt to the construction of the string Rindler modes
different from those of Sec.\ 3 is presented.
The final section (Sec.\ 6) is devoted to a summary of the remaining
problems. This paper contains four Appendices. In three of them (A,
B and D), various formulas used in the text are derived. In Appendix
C, we present the proof of the BRST invariance of the Rindler vacuum
defined in Sec.\ 4.

\section{Free string field theory on Minkowski space-time}
\label{sec:SFT_M}

Before starting the construction of SFT on Rindler space-time, we
shall recapitulate the elements of the free SFT covariantly quantized
on Minkowski space-time.
In this paper we shall confine ourselves to open bosonic SFT.

First, the string field $\Phi$ is a real ($\Phi^\dagger=\Phi$)
functional of the space-time string coordinate $X^\mu(\sigma)$ as well
as the ghost and anti-ghost coordinates, $c(\sigma)$ and
$\ac(\sigma)$ ($0\le\sigma\le\pi$):
\begin{equation}
\Phi=\Phi\left[X^\mu(\sigma),c(\sigma), \ac(\sigma)\right] .
\label{eq:Phi_M}
\end{equation}
In the following we shall consider the SFT which is gauge-fixed in the
Siegel gauge \cite{Siegel}, and therefore
$\ac(\sigma)$ in (\ref{eq:Phi_M}) (and in all the equations below)
should be understood to be free from the zero-mode.
Then the BRST invariant action for $\Phi$ reads
\cite{FreeSFT,WittenSFT,HIKKOopen}
\begin{eqnarray}
S=-\Half\int\calD X^\mu(\sigma)\,\calD c(\sigma)\,\calD \ac(\sigma)
\,\Phi L\Phi\ ,
\label{eq:S_M}
\end{eqnarray}
and the corresponding equation of motion is given by
\begin{equation}
L\Phi=0\ .
\label{eq:LPhi=0}
\end{equation}
In eqs.\ (\ref{eq:S_M}) and (\ref{eq:LPhi=0}), $L$ is the kinetic
operator,
\begin{eqnarray}
&&L=\pi\int_0^\pi d\sigma\left\{
-\eta^{\mu\nu}\hanbi{}{X^\mu}\hanbi{}{X^\nu}
+\eta_{\mu\nu}{X'^\mu}{X'^\nu}
+\mbox{(ghost part)}
\right\}\nn \\
&&\phantom{L}
= -\eta^{\mu\nu}\henbi{}{x^\mu}\henbi{}{x^\nu}
+2\sum_{n=1}^{\infty}\eta^{\mu\nu}\left\{
-\henbi{}{x_n^\mu}\henbi{}{x_n^\nu}
+\left(\frac{n\pi}{2}\right)^2 x_n^\mu x_n^\nu\right\}
+(\mbox{ghost part})\nn\\
&&\phantom{L}
=-\eta^{\mu\nu}\henbi{}{x^\mu}\henbi{}{x^\nu}
+[\mbox{$(\mbox{mass})^2$-operator}]\ ,
\label{eq:L_M}
\end{eqnarray}
where $\eta_{\mu\nu}=\eta^{\mu\nu}={\rm diag}(-1,1,\cdots,1)$ is the
flat Minkowski metric, and we define the Fourier expansion coefficient
$(x^\mu, x_n^\mu)$ of $X^\mu(\sigma)$ by
\begin{eqnarray}
&&X^\mu(\sigma)=x^\mu+\sum_{n=1}^{\infty}x^\mu_n\cos{n\sigma}
\label{eq:X_mode}\ ,\\
&&\hanbi{}{X^\mu(\sigma)}=\frac{1}{\pi}\left\{
\henbi{}{x^\mu} + 2\sum_{n=1}^{\infty}\cos(n\sigma)\henbi{}{x^\mu_n}
\right\}\ .
\label{eq:d/dX_mode}
\end{eqnarray}
The coordinate $x^\mu$ is the same as $x^\mu_{\rm CM}$ in Sec.\ 1.
We omit the subscript CM for simplicity.

The Minkowski quantization of this SFT system is carried out
by taking the CM coordinate $x^0$ of $X^{\mu=0}(\sigma)$ as the time
coordinate. Namely, defining the canonical momentum $\Pi_{\rm M}$
conjugate to $\Phi$ by
\begin{equation}
\Pi_{\rm M}\left[X^\mu(\sigma),c(\sigma), \ac(\sigma)\right]
\equiv \hanbi{S}{(\p\Phi/\p x^0)}
=\henbi{}{x^0}\Phi\left[X^\mu(\sigma),c(\sigma), \ac(\sigma)\right]\ ,
\end{equation}
we impose the equal $x^0$ canonical commutation relation,
\begin{eqnarray}
&&\Bigl[\Phi[X, c, \ac\,],
\Pi_{\rm M}[\til{X},\til{c},\til{\ac}\,]\Bigr]
\Big\vert_{x^0=\til{x}^0}\nn\\
&&\quad
=i\prod_\sigma \widehat{\delta}\left(X(\sigma)-\til{X}(\sigma)\right)
\delta\left(c(\sigma)-\til{c}(\sigma)\right)
\delta\left(\ac(\sigma)-\til{\ac}(\sigma)\right)\ ,
\label{eq:CCR_M}
\end{eqnarray}
where the hat on $\widehat{\delta}(X-\til{X})$ implies that the delta
function for the time variable, $\delta(x^0-\til{x}^0)$, is missing.

Then we expand the string field $\Phi$ in terms of the complete set of
the Minkowski modes $\left\{U_{k\{M,N\}A},U^*_{k\{M,N\}A}\right\}$:
\begin{equation}
\Phi=\int_{-\infty}^\infty dk \sum_{M_n,N_n=0}^\infty\sum_A
\left(U_{k\{M,N\}A} a_{k\{M,N\}}^A
+a_{k\{M,N\}}^{A\dagger} U^*_{k\{M,N\}A}\right)\ ,
\label{eq:Phi=Ua}
\end{equation}
which manifestly respects the hermiticity of $\Phi$.
$U_{k\{M,N\}A}$ is the normalized positive frequency solution (with
respect to the Minkowski time $x^0$) of the wave equation,
\begin{equation}
L U_{k\{M,N\}A}=0\ ,
\end{equation}
and the complex conjugate $U^*_{k\{M,N\}A}$ is the negative
frequency one.
The composition of our Minkowski mode $U_{k\{M,N\}A}$ is rather
unconventional, however it is convenient for the analysis of the
Rindler quantization in later sections.
The meaning of the indices of $U_{k\{M,N\}A}$ is as follows:
$k$ is the momentum conjugate to the CM coordinate $x^1$, $\{M,N\}$
denotes the set of the level numbers $(M_n,N_n)$ of the oscillators
$(x^0_n,x^1_n)$ ($n=1,2.\cdots$), and the index
$A$ represents collectively the transverse momentum $\bm{k}_\bot$ and
the level numbers of the $(X^\bot,c,\ac)$ oscillators ($X^\bot$
denotes $X^\mu$ with $\mu=2, \cdots, 25$).
Therefore, the symbol $\sum_A$ implies the integrations and summations
over the momentum variables and the level numbers contained in $A$.

Corresponding to the three sets of the quantum numbers $k$, $\{M,N\}$
and $A$, the Minkowski mode $U_{k\{M,N\}A}$ is given as the product
of three components:
\begin{equation}
U_{k\{M,N\}A}=U_k(x^0,x^1)\cdot\Psi_{\{M,N\}}(x^0_n,x^1_n)
\cdot\phi_A\ ,
\label{eq:U_M}
\end{equation}
where
\begin{eqnarray}
&&U_k(x^0,x^1)=\frac{1}{\sqrt{2\omega_k\cdot 2\pi}}
\exp\left(-i\omega_k x^0+ikx^1\right)\ ,
\label{eq:U_k}\\
&&\Psi_{\{M,N\}}(x^0_n,x^1_n)=
\prod_{n=1}^\infty \Psi_{M_n,N_n}(x^0_n,x^1_n)\ ,
\label{eq:Psi_MN}\\
&&\Psi_{M_n,N_n}(x^0_n,x^1_n)=i^{M_n}
\psi^{(n)}_{M_n}(ix^0_n)\,\psi^{(n)}_{N_n}(x^1_n)\ ,
\label{eq:Psi^n_MN}
\end{eqnarray}
and $\phi_A$ is the wave function for the transverse and the
ghost coordinates $(X^\bot,c,\ac)$. Various quantities appearing in
(\ref{eq:U_k}) and (\ref{eq:Psi_MN}) are as follows:
\begin{eqnarray}
&&\omega_k=\sqrt{k^2+\mu^2}, \label{eq:omega_k}\\
&&\mu^2={\bm{k}_\bot}^2+m^2,\\
&&m^2=m^2\left(\{M,N\},A\right)=
2\pi\sum_{n=1}^\infty n\left(M_n + N_n\right) + m_A^2 ,\\
&&\psi^{(n)}_M(x)=\sqrt{\frac{1}{M!}\sqrt{\frac{n}{2}}}
e^{-n\pi x^2/4}H_M\left(\sqrt{n\pi}x\right),
\label{eq:psi^n}
\end{eqnarray}
where $m_A^2$ is the (mass)$^2$ contributing from the transverse and
ghost oscillators ($m_A^2$ also contains the intercept term $-2\pi$),
and $H_M(x)$ is the Hermite polynomial defined by
$H_M(x)=e^{x^2/2}\left(-d/dx\right)^M e^{-x^2/2}$.
Due to the factor $i^{M_n}$ in eq.\ (\ref{eq:Psi^n_MN}),
$\Psi^{(n)}_{M_n,N_n}$ is real when $x^0_n$ is real.
In the above definitions we have used some abbreviations for the
indices for the sake of simplicity.
First, although $U_k(x^0,x^1)$ depends on $\{M,N\}$ and $A$ through
$\omega_k$ and therefore should carry them as its index,
we have omitted this dependence.
Second, we have omitted the superscript $(n)$ which
$\Psi_{M_n,N_n}$ (\ref{eq:Psi^n_MN}) should carry like
$\psi^{(n)}_{M}$ (\ref{eq:psi^n}) since it can be specified by the
subscript $n$ of $(M_n, N_n)$.

For the normalization of the Minkowski modes, we define
the Minkowski inner product for the string wave functions $\Phi_1$ and
$\Phi_2$ by
\begin{eqnarray}
&&\left(\Phi_1,\Phi_2\right)=
i\int \frac{\cald X^\mu(\sigma) \cald c(\sigma) \cald \ac(\sigma)}
{dx^0}\,
\Phi^*_1\sayuu{\henbi{}{x^0}}\Phi_2\nn \\
&&\phantom{\left(\Phi_1,\Phi_2\right)}
=i\prod_{k=1}^{D-1}\int_{-\infty}^{\infty} dx^k
\prod_{n=1}^\infty\int_{-\infty}^{\infty}dx^k_n
\int_{-\infty i}^{\infty i}i\,dx^0_n
\int\cald c(\sigma) \cald \ac(\sigma)\nn\\
&&\hspace*{4cm}
\times\Phi^*_1(x_n^0)\sayuu{\henbi{}{x^0}}\Phi_2(x_n^0),
\label{eq:innprod_M}
\end{eqnarray}
where we have made explicit only the $x^0_n$-dependence of the wave
functions. The integration in (\ref{eq:innprod_M}) is carried out over
the string coordinates
$\left(X^\mu(\sigma),c(\sigma),\ac(\sigma)\right)$
with a fixed Minkowski CM time coordinate $x^0$.
If both $\Phi_1$ and $\Phi_2$ satisfy the wave equation
$L\Phi_{1,2}=0$, the inner product (\ref{eq:innprod_M}) is
conserved, namely, it is independent of the choice of $x^0$.

We have to explain the prescription concerning the variables $x_n^0$
in the inner product (\ref{eq:innprod_M}).
First, the complex conjugation operation on
$\Phi_1$ in eq.\ (\ref{eq:innprod_M}) should be carried out by
regarding $x_n^0$ as real variables. Then, the integrations over
$x_n^0$ are carried out in the pure-imaginary direction.
This prescription is due to the fact in the first quantized string
theory that $x_n^0$ is a negative-norm harmonic oscillator with
pure-imaginary eigenvalues, and hence the completeness
relation for the eigenstates of $x_n^0$ reads
$\int_{-\infty i}^{\infty i}i dx_n^0 \ket{-x_n^0}\bra{x_n^0}
=1\kern-1ex 1$ \cite{AFIO}. This is consistent with the inner product
using the indefinite metric creation/annihilation operators in the
first quantized string theory.
The complex conjugation operation in all other places than the inner
product should also be done by regarding $x_n^0$ as real variables.

Using the above inner product (\ref{eq:innprod_M}), the Minkowski
modes $\{U,U^*\}$ are normalized as follows:\footnote{
The following two formulas are useful:
$(\Phi_1,\Phi_2)=(\Phi_2,\Phi_1)^*=-(-)^{|\Phi_1|}(\Phi_2^*,\Phi_1^*)$
and
$\eta^*_{AB}=\eta_{BA}$ where $|\Phi|=0$ ($1$) if $\Phi$ is
Grassmann-even (odd). We assume that the measure
$\int\!\calD c\,\calD \ac$ is hermitian.
}
\begin{eqnarray}
&&\left(U_{k\{M,N\}A},
U_{k'\{M',N'\}A'}\right)
=\delta(k-k')\,\delta_{\{M,N\},\{M',N'\}}
(-)^{\sum_n{M_n}}\eta_{AA'}\ ,\nn\\
&&\left(U^*_{k\{M,N\}A},
U^*_{k'\{M',N'\}A'}\right)
=-(-)^{|A|}\delta(k-k')\,\delta_{\{M,N\},\{M',N'\}}
(-)^{\sum_n{M_n}}\eta_{A'A}\ ,
\label{eq:(U,U)_M}\\
&&\left(U_{k\{M,N\}A},
U^*_{k'\{M',N'\}A'}\right)=0\ ,\nn
\end{eqnarray}
where we have used the abbreviation
$\delta_{\{M,N\},\{M',N'\}}\equiv\prod_n\delta_{M_n,M'_n}\delta_{N_n,N'_n}$,
and $\eta_{AA'}$ is the inner product of the transverse wave
functions,
\begin{equation}
\eta_{AA'}=\left(\phi_A,\phi_{A'}\right)\equiv
\int\calD X^\bot \calD c\, \calD \ac\, \phi^*_A \phi_{A'}\ ,
\label{eq:eta_AA}
\end{equation}
and $|A|$ in $(-)^{|A|}$ is defined by
\begin{equation}
|A|=\cases{ 0  &if $\phi_A$ is Grassmann-even\ ,\cr
            1  &if $\phi_A$ is Grassmann-odd\ .}
\label{eq:|A|}
\end{equation}
Eq.\ (\ref{eq:(U,U)_M}) is the product of the inner products of the
factor wave functions $U_k$, $\Psi_{M_n,N_n}$ and $\phi_A$.
The inner product for $U_k(x^0,x^1)$,
$(U_k,U_{k'})\equiv i\int_{-\infty}^\infty
dx^1 U_k^*(x^0,x^1)(\sayuu{\p}/\p x^0)U_{k'}(x^0,x^1)$,
is given by
\begin{eqnarray}
&&\left(U_k,U_{k'}\right)=-\left(U_k^*,U_{k'}^*\right)
=\delta(k-k')\ ,\nn\\
&&\left(U_k,U_{k'}^*\right)=\left(U_k^*,U_{k'}\right)=0\ ,
\label{eq:(U_k,U_k)}
\end{eqnarray}
where $U_k$ and $U_{k'}$ should have a common $\mu^2$
of eq.\ (\ref{eq:omega_k}).
As for the inner product of $\Psi_{M_n,N_n}$ defined by
\begin{equation}(\Psi,\widetilde\Psi)=
\int_{-\infty}^{\infty}dx^1_n
\int_{-\infty i}^{\infty i}i\,dx^0_n
\Psi^*(x^0_n,x^1_n)\widetilde\Psi(x^0_n,x^1_n)\ ,
\label{eq:innprod_Psi}
\end{equation}
we have
\begin{equation}
\left(\Psi_{M_n,N_n},\Psi_{M'_n,N'_n}\right)
=\delta_{M_n,M'_n}\delta_{N_n,N'_n}(-)^{M_n} .
\label{eq:(Psi_MN,Psi_MN)}
\end{equation}
In eq.\ (\ref{eq:innprod_Psi}), the meaning of the complex conjugation
is the same as explained below eq.\ (\ref{eq:innprod_M}).

Let us return to eq.\ (\ref{eq:Phi=Ua}).
Using the orthonormality (\ref{eq:(U,U)_M}),
$a_{k\{M,N\}}^A$ and $a_{k\{M,N\}}^{A\dagger}$ are expressed in
terms of $\Phi$ as
\begin{eqnarray}
&&a_{k\{M,N\}}^A=(-)^{\sum_n{M_n}}\sum_{A'}\eta^{AA'}
\left(U_{k\{M,N\}A'},\Phi\right)\ ,\nn\\
&&a_{k\{M,N\}}^{A\dagger}=-(-)^{\sum_n{M_n}}\sum_{A'}
\left(U^*_{k\{M,N\}A'},\Phi\right)\eta^{A'A}\ .
\label{eq:a=(U,Phi)}
\end{eqnarray}
The original string field $\Phi$ is Grassmann-even, and hence
$a_{k\{M,N\}}^A$ is Grassmann-even (odd) if $\phi_A$ and hence
$U_{k\{M,N\}A}$ are Grassmann-even (odd).
Then using the canonical commutation relations (\ref{eq:CCR_M}),
we can show the following (anti-)commutation relation between
the creation and annihilation operators:
\begin{eqnarray}
\CR{a_{k\{M,N\}}^A,a_{k'\{M',N'\}}^{A'\dagger}}
\!\!&\equiv&\!\!
a_{k\{M,N\}}^A a_{k'\{M',N'\}}^{A'}
-(-)^{|A||A'|}a_{k'\{M',N'\}}^{A'} a_{k\{M,N\}}^A\nn\\
\!\!&=&\!\!
\delta(k-k')\,\delta_{\{M,N\},\{M',N'\}}(-)^{\sum_n{M_n}}\eta^{AA'}\ ,
\end{eqnarray}
where $\eta^{AA'}$ is the inverse of $\eta_{AA'}$:
\begin{equation}
\sum_{B}\eta_{AB}\eta^{BA'}=\sum_{B}\eta^{AB}\eta_{BA'}
=\delta_A^{A'}\ .
\label{eq:eta^AB}
\end{equation}

Finally, the Minkowski vacuum state $\mket{0}$ is defined as usual by
\begin{equation}
a_{k\{M,N\}}^A\mket{0}=0\ .
\label{eq:vac_M}
\end{equation}
This is the lowest eigenstate of the Minkowski
Hamiltonian $H_{\rm M}$,
\begin{equation}
H_{\rm M}=\int_{-\infty}^\infty\!
dk \sum_{\{M,N\}}\sum_{A,B}\omega_k
\cdot a_{k\{M,N\}}^{A\dagger}a_{k\{M,N\}}^B
\cdot (-)^{\sum_n M_n}\eta_{AB}\ ,
\label{eq:H_M}
\end{equation}
which is the generator of the $x^0$ translation:
\begin{equation}
\left[H_{\rm M}, \Phi\right]=-i\henbi{}{x^0}\Phi\ .
\end{equation}
Recall that $\omega_k$ depends also on $\{M, N\}$ and $A$.

\section{String Rindler modes}
\label{sec:SRM}

As a first step toward the Rindler quantization of SFT, we shall
construct the string Rindler modes, namely,
a complete set of solution
$u_\Omega^{(\sigma)}$ of the string field equation in the Rindler
wedge $\sigma$ ($\sigma=\pm$) having the Rindler energy $\Omega$
($\Omega>0$).
Since the string field $\Phi$ is a space-time scalar, the Rindler
wave equation for $u_\Omega^{(\sigma)}$ is given simply by
\begin{equation}
Lu_\Omega^{(\sigma)}=0\  ,
\label{eq:Lu=0}
\end{equation}
using the same kinetic operator $L$ as given by (\ref{eq:L_M}) in the
Minkowski quantization (c.f., Appendix \ref{app:innprod_R}).
The Rindler mode $u_\Omega^{(\sigma)}$ is requested to satisfy
the following three conditions:

\begin{enumerate}

\item[(I)] $u_\Omega^{(\sigma)}$ is the positive frequency mode with
energy $\Omega$ with respect to the Rindler time $\eta_0$
(\ref{eq:eta_0}). Namely, we have
$u_\Omega^{(\sigma)}\propto\exp\left(-i\sigma\Omega\eta_0\right)$,
and hence
\begin{equation}
\henbi{}{\eta_0}u_\Omega^{(\sigma)}=-i\sigma\Omega
u_\Omega^{(\sigma)}\ .
\label{eq:du/deta_0}
\end{equation}

\item[(II)] $u_\Omega^{(\sigma)}$ is normalized (with respect to the
Rindler inner product).

\item[(III)] $u_\Omega^{(\pm)}$ vanishes in the $\mp$-wedge
(Wedge condition).

\end{enumerate}

We should add a few comments on these conditions. First,
as stated in Sec.\ 1, we take as the Rindler time for quantization the
CM coordinate $\eta_0$ given by (\ref{eq:eta_0}).
The reason why $\sigma$ is multiplied on the RHS of eq.\
(\ref{eq:du/deta_0}) is that we go to the past as $\eta_0$ is
increased in the negative Rindler wedge.
Second, we can (formally) define the conserved inner product
$(*,*)_{\rm R}$
for the Rindler quantization as given by eq.\ (\ref{eq:innprod_R}) in
Appendix \ref{app:innprod_R}, and the inner product for the condition
(II) should be this one. However, we assume throughout this paper that
the Rindler inner product equals the Minkowski inner product
(\ref{eq:innprod_M}): this assumption is fairly reasonable as
is explained in Appendix \ref{app:innprod_R}.
Among the above three conditions, the last one (III), which we call
{\em wedge condition} hereafter, is the most
difficult and obscure condition in the SFT case as explained
in Sec.\ 1. Note that the wedge condition of the Rindler modes
determines the way of dividing the Hilbert space of states into those
of the $+$ and $-$ wedges.

At present we are omitting possible other indices (quantum
numbers) that the Rindler mode $u_\Omega^{(\sigma)}$ should carry.
These quantum numbers will become clear in the course of the
construction of the Rindler modes.

\subsection{Condition (I)}

In this paper, we shall not attempt constructing the string Rindler
modes directly as a function of the string Rindler coordinate
$\left(\xi(\sigma),\eta(\sigma),X^\bot(\sigma)\right)$.
Rather, we shall express the Rindler modes as a linear combination of
the Minkowski modes of Sec.\ 2. This is possible
since the kinetic operator $L$ for the Rindler wave equation
(\ref{eq:Lu=0}) is actually the same as for the Minkowski modes.
Taking as the quantum numbers for the transverse and the ghost
coordinates the same $A$ as for the Minkowski modes,
let us express $u_{\Omega,A}^{(\sigma)}$ (carrying the index $A$) as
\begin{equation}
u_{\Omega,A}^{(\sigma)}=
\int_{-\infty}^\infty\! dk \sum_{M,N}\left(
\alpha_{\Omega}^{k\{M,N\}(\sigma)} U_{k}
+\beta_{\Omega}^{k\{M,N\}(\sigma)} U_{k}^*\right)
\Psi_{\{M,N\}}\cdot\phi_A\ ,
\label{eq:Expand_u}
\end{equation}
where $U_{k}$ and $\Psi_{\{M,N\}}$ are given by eqs.\ (\ref{eq:U_k})
and (\ref{eq:Psi_MN}), and $\alpha_{\Omega}^{k\{M,N\}(\sigma)}$ and
$\beta_{\Omega}^{k\{M,N\}(\sigma)}$ are the coefficients to be
determined below.
$u_{\Omega,A}^{(\sigma)}$ given by (\ref{eq:Expand_u}) obviously
satisfies the wave equation (\ref{eq:Lu=0}), and we next impose the
condition (\ref{eq:du/deta_0}).
For this purpose, we note first that
\begin{eqnarray}
&&\henbi{}{\eta_0}=\int_0^\pi d\sigma\hanbi{}{\eta(\sigma)}
=\int_0^\pi d\sigma \left(X^0(\sigma)\hanbi{}{X^1(\sigma)}
+ X^1(\sigma)\hanbi{}{X^0(\sigma)}\right)\nn\\
&&\phantom{\henbi{}{\eta_0}}
=x^0\henbi{}{x^1}+x^1\henbi{}{x^0}+
\sum_{n=1}^\infty
\left(x_n^0\henbi{}{x_n^1}+x_n^1\henbi{}{x_n^0}\right)\ ,
\label{eq:d/deta_0}
\end{eqnarray}
and then use the following properties for $U_k$ (\ref{eq:U_k}) and
$\Psi_{M_n,N_n}$ (\ref{eq:Psi^n_MN}):
\begin{eqnarray}
&&\left(x^0\henbi{}{x^1}+x^1\henbi{}{x^0}\right)U_k(x^0,x^1)
= -\sqrt{\omega_k}\henbi{}{k}\left(\sqrt{\omega_k}U_k(x^0,x^1)
\right)\ ,
\label{eq:dU/deta_0}\\
&&\left(x_n^0\henbi{}{x_n^1}+x_n^1\henbi{}{x_n^0}\right)
\Psi_{M_n,N_n}(x^0_n,x^1_n)\nn\\
&&\qquad
=-\sqrt{M_n(N_n+1)}\Psi_{M_n-1,N_n+1}
-\sqrt{(M_n+1)N_n}\Psi_{M_n+1,N_n-1}\ ,
\label{eq:dPsi_n/deta_0}
\end{eqnarray}
where eq.\ (\ref{eq:dPsi_n/deta_0}) is a consequence of the two
formulas for the Hermite polynomials:
\begin{eqnarray}
&&\frac{d}{dx}H_n(x)=nH_{n-1}(x)\ ,\nn\\
&&xH_n(x)=H_{n+1}(x)+nH_{n-1}(x)\ .
\end{eqnarray}
Making the $k$-integration by parts for the term arising from eq.\
(\ref{eq:dU/deta_0}),\footnote{
See Sec.\ \ref{sec:BC_u-I} for the validity of discarding the
surface term.}
the condition (\ref{eq:du/deta_0}) is reduced to the following
equation for the coefficient $\alpha_{\Omega}^{k\{M,N\}(\sigma)}$
and exactly the same one for $\beta_{\Omega}^{k\{M,N\}(\sigma)}$:
\begin{eqnarray}
&&\left(\sqrt{\omega_k}\henbi{}{k}\sqrt{\omega_k}
+ i\sigma\Omega\right)\alpha_{\Omega}^{k\{M,N\}(\sigma)}\nn\\
&&
=\sum_{n=1}^\infty \left(
\sqrt{(M_n+1)N_n}\,\alpha_{\Omega}^{k\{M+1_n,N-1_n\}(\sigma)}
+\sqrt{M_n(N_n+1)}\,\alpha_{\Omega}^{k\{M-1_n,N+1_n\}(\sigma)}
\right)\ ,
\label{eq:Eq_for_alpha}
\end{eqnarray}
where the meaning of $M\pm 1_n$ is
\begin{equation}
(M\pm 1_n)_m=\cases{M_n\pm 1 & $(m=n)$\cr
                    M_m      & $(m\ne n)$\ .
}
\label{eq:Mpm1n}
\end{equation}
In deriving eq.\ (\ref{eq:Eq_for_alpha}) we have used the fact that
$U_k(x^0,x^1)$ depends on $\{M,N\}$ only through the combination
$\sum_n n\left(M_n+N_n\right)$, and hence the $U_k$s associated with
the three terms of eq.\ (\ref{eq:Eq_for_alpha}) in the original
eq.\ (\ref{eq:du/deta_0}) are common.

To solve the differential-recursion equation (\ref{eq:Eq_for_alpha}),
we should note that
\begin{equation}
T_n\equiv M_n + N_n\ ,
\label{eq:T_n}
\end{equation}
is common for the three terms of (\ref{eq:Eq_for_alpha}) and hence
we can consider the solution with a fixed $T_n$.
Then, eq.\ (\ref{eq:Eq_for_alpha_0}) is
solved by the separation of variables. Namely,
we assume that the dependences of
$\alpha_{\Omega}^{k\{M,N\}(\sigma)}$ on $k$ and $M_n$
$(n=1,2,\cdots)$ are factorized,
\begin{equation}
\alpha_{\Omega}^{k\{M,N\}(\sigma)}
=\alpha_0(k)\prod_{n=1}^\infty \alpha_n(M_n)\ ,
\quad (M_n+N_n=T_n)
\label{eq:fac_alpha}
\end{equation}
where the dependence on $(\Omega,\sigma)$ is omitted on the RHS for
the sake of simplicity.
Then, since $\omega_k$ depends on $\{M,N\}$ only through $\{T\}$,
\begin{equation}
\omega_k=\sqrt{k^2+\bm{k}_\bot^2 + 2\pi\sum_n nT_n + m_A^2}\ ,
\end{equation}
$\alpha_0$ and $\alpha_n$ should satisfy the following equations:
\begin{eqnarray}
&&\left(\sqrt{\omega_k}\henbi{}{k}\sqrt{\omega_k}
+ i\sigma\Omega + \sum_{n=1}^\infty\lambda_n \right)
\alpha_0(k)=0\ ,
\label{eq:Eq_for_alpha_0}\\[7pt]
&&\sqrt{(M_n+1)(T_n-M_n)}\,\alpha_n(M_n+1)
+\sqrt{M_n(T_n-M_n+1)}\,\alpha_n(M_n-1)\nn\\
&&\hspace*{3cm}
+\lambda_n\alpha_n(M_n)=0\ ,
\label{eq:Eq_for_alpha_n}
\end{eqnarray}
where the constants $\lambda_n$ ($n=1,2,\cdots$) are to be determined
as eigenvalues of the recursion equation (\ref{eq:Eq_for_alpha_n}).
Our solution is now characterized by the set of quantum numbers
$\{T,\lambda\}$.

The solution to the differential equation (\ref{eq:Eq_for_alpha_0}) is
given by
\begin{equation}
\alpha_0(k)=\frac{1}{\sqrt{2\pi\omega_k}}
\left(\frac{\omega_k+k}{\omega_k-k}\right)^{
-\left(i\sigma\Omega+\sum_n\lambda_n\right)/2}\ ,
\label{eq:alpha_0}
\end{equation}
up to an overall constant.
As for eq.\ (\ref{eq:Eq_for_alpha_n}), the solution $\alpha_n(M_n)$ and
the allowed values of $\lambda_n$ and $T_n$ are found from the
angular momentum formula in three dimensions:
\begin{equation}
2 \widehat{J}_x\ket{j,j_z;z}=
\sqrt{(j-j_z)(j+j_z+1)}\ket{j,j_z + 1;z}
+\sqrt{(j+j_z)(j-j_z+1)}\ket{j,j_z-1;z} ,
\label{eq:Jx}
\end{equation}
where the ket $\ket{j,j_z;z}$ is the eigenstate of
$\widehat{\bm{J}}^2$ and $\widehat{J}_z$ with eigenvalues $j(j+1)$ and
$j_z$, respectively. In fact, by taking the inner product between
eq.\ (\ref{eq:Jx}) and the bra-state $\bra{j,j_x;x}$, which is the
eigenstate of $\widehat{J}_x$ with eigenvalue $j_x$, we find the
following correspondence:
\begin{eqnarray}
&&T=2j ,\quad M=j+j_z ,\quad N=j-j_z ,\\
&&\lambda=-2j_x ,\\
&&\alpha_{T,\lambda}(M)= \VEV{
\frac{T}{2},-\frac{\lambda}{2};x\Bigg|
\frac{T}{2},M-\frac{T}{2};z} , \label{eq:alpha_Tlambda}
\end{eqnarray}
where we have omitted the subscript $n$ and attached the index
$(T,\lambda)$ to $\alpha(M)$.
Therefore the allowed values of $(T_n,\lambda_n)$ are
\begin{eqnarray}
&&T_n=0,1,2,\cdots \label{eq:allowedT_n}\\
&&\lambda_n=T_n, T_n\!-\!2, T_n\!-\!4, \cdots, -T_n\!+\!2, -T_n\ .
\label{eq:allowedlambda_n}
\end{eqnarray}
$\alpha_{T,\lambda}(M)$ of eq.\ (\ref{eq:alpha_Tlambda}) can be chosen
to be real for all $M$,\footnote{
Once we choose $\alpha_{T,\lambda}(M=0)$ to be real,
$\alpha_{T,\lambda}(M)$ ($M\ge 1$) determined from the recursion
relation (\ref{eq:Eq_for_alpha_n}) are all real.}
and they are orthonormal with respect to
$\lambda$ for a common $T$:
\begin{equation}
\sum_{M=0}^T \alpha_{T,\lambda}(M)\alpha_{T,\lambda'}(M)
=\delta_{\lambda,\lambda'}\ .
\label{eq:sum_alpha-alpha}
\end{equation}
Another useful formula for $\alpha_{T,\lambda}(M)$ is
\begin{equation}
\alpha_{T,-\lambda}(M)=(-)^M \alpha_{T,\lambda}(M)\ .
\label{eq:alpha_T(-lambda)}
\end{equation}

Recalling that $\beta_{\Omega}^{k\{M,N\}(\sigma)}$ should satisfy
exactly the same equation (\ref{eq:Eq_for_alpha}) as for
$\alpha_{\Omega}^{k\{M,N\}(\sigma)}$,
the string Rindler mode satisfying eq.\ (\ref{eq:du/deta_0}) and
labeled by $\{T,\lambda\}$ is given as:
\begin{equation}
u_{\Omega\{T,\lambda\}A}^{(\sigma)}=
\int_{-\infty}^\infty\! \frac{dk}{\sqrt{2\pi\omega_k}}
\left(\frac{\omega_k+k}{\omega_k-k}\right)^{\!
-\left(i\sigma\Omega+\sum_n\lambda_n\right)/2}\kern-4pt
\left(\alpha_{\Omega\{T,\lambda\}}^{(\sigma)} U_k
+ \beta_{\Omega\{T,\lambda\}}^{(\sigma)} U_k^*\right)
\Phi_{\{T,\lambda\}}
\cdot\phi_A ,
\label{eq:u_Tlambda}
\end{equation}
where $\alpha_{\Omega\{T,\lambda\}}^{(\sigma)}$ and
$\beta_{\Omega\{T,\lambda\}}^{(\sigma)}$ are constants undetermined
at this stage, and $\Phi_{\{T,\lambda\}}$ is given by
\begin{eqnarray}
&&\Phi_{\{T,\lambda\}}=\prod_{n=1}^\infty\Phi_{T_n,\lambda_n}\ ,
\label{eq:Phi_Tlambda}\\
&&\Phi_{T_n,\lambda_n}=\sum_{M_n=0}^{T_n}\alpha_{T_n,\lambda_n}(M_n)
\Psi_{M_n,T_n-M_n}\ ,
\label{eq:Phi_T_nlambda_n}
\end{eqnarray}
with $\Psi_{M_n,N_n}$ of eq.\ (\ref{eq:Psi^n_MN}).

Looking back the above derivation of
$u_{\Omega\{T,\lambda\}A}^{(\sigma)}$,
we find that the following two equations hold:
\begin{eqnarray}
&&\left(x^0\henbi{}{x^1}+x^1\henbi{}{x^0}
+ i\sigma\Omega +\sum_n \lambda_n\right)
u_{\Omega\{T,\lambda\}A}^{(\sigma)}=0\ ,
\label{eq:complex_energy}\\
&&\left(x_n^0\henbi{}{x_n^1}+x_n^1\henbi{}{x_n^0}-\lambda_n
\right)\Phi_{T_n,\lambda_n}=0\ .
\label{eq:alpha_nPsi_MnNn}
\end{eqnarray}
In particular, since we have
$\p/\p\oeta=x^0\left(\p/\p x^1\right)+x^1\left(\p/\p x^0\right)$
for the string Rindler time $\oeta$ of eq.\ (\ref{eq:oeta_oxi}) in the
naive approach mentioned in Sec.\ 1,
eq.\ (\ref{eq:complex_energy}) implies that
our $u_{\Omega\{T,\lambda\}A}^{(\sigma)}$ carries the complex energy
$\Omega-i\sigma\sum_n\lambda_n$ with respect to the $\oeta$
time.

The Rindler modes $u_{\Omega\{T,\lambda\}A}^{(\sigma)}$
(\ref{eq:u_Tlambda}) is specified by the the index $\{T,\lambda\}$.
However, the quantum number $\{T,\lambda\}$ may not be a good one when
we take into account the wedge condition (condition (III)).
Then we would have to consider a more general wave function,
\begin{equation}
u_{\Omega,Z,A}^{(\sigma)}=\sum_{\{T,\lambda\}}
\int_{-\infty}^\infty\! \frac{dk}{\sqrt{2\pi\omega_k}}
\left(\frac{\omega_k+k}{\omega_k-k}\right)^{\!
-\left(i\sigma\Omega+\sum_n\lambda_n\right)/2}\kern-4pt
\left(\alpha_{\Omega,Z,\{T,\lambda\}}^{(\sigma)} U_k
+ \beta_{\Omega,Z,\{T,\lambda\}}^{(\sigma)} U_k^*\right)
\Phi_{\{T,\lambda\}}
\cdot\phi_A ,
\label{eq:u_Z}
\end{equation}
obtained by summing over $\{T,\lambda\}$ and labeled by a new quantum
number $Z$. We shall try to construct this type of Rindler modes in
Sec.\ 5.

\subsection{Normalization of the Rindler modes}

Next we shall impose the condition that the Rindler modes
$u_{\Omega\{T,\lambda\}A}^{(\sigma)}$ as given by
(\ref{eq:u_Tlambda}) be properly normalized (condition (II)).
As stated at the beginning of this section, we shall employ the
Minkowski inner product, which we assume to be equal to the Rindler
one.
For calculating the inner product between two
$u_{\Omega\{T,\lambda\}A}^{(\sigma)}$, note first that
$\Phi_{\{T,\lambda\}}$ satisfies the following normalization:
\begin{eqnarray}
\left(\Phi_{\{T,\lambda\}},\Phi_{\{T',\lambda'\}}\right)
=\dTlambda\equiv
\prod_{n=1}^\infty\delta_{T_n,T'_n}\delta_{\lambda_n+\lambda'_n,0}\ ,
\label{eq:(Phi,Phi)}
\end{eqnarray}
which follows from eqs.\
(\ref{eq:(Psi_MN,Psi_MN)}), (\ref{eq:sum_alpha-alpha}),
(\ref{eq:alpha_T(-lambda)}) and (\ref{eq:Phi_Tlambda}).
The non-diagonal form of (\ref{eq:(Phi,Phi)}) with respect to
$\lambda_n$ originates from the fact that $x^0_n$ is a negative norm
oscillator.
Then using eqs.\ (\ref{eq:eta_AA}), (\ref{eq:(U_k,U_k)}) and
(\ref{eq:(Phi,Phi)}), we find that the inner product between two
$u_{\Omega\{T,\lambda\}A}^{(\sigma)}$ is given by
\begin{eqnarray}
&&\left(u_{\Omega\{T,\lambda\}A}^{(\sigma)},
u_{\Omega'\{T',\lambda'\}A'}^{(\sigma')} \right)
=\int_{-\infty}^\infty\frac{dk}{2\pi\omega_k}
\left(\frac{\omega_k+k}{\omega_k-k}\right)^{\!
i\left(\sigma\Omega-\sigma'\Omega'\right)/2
-\sum_n\left(\lambda_n+\lambda'_n\right)/2}\nn\\
&&\qquad\qquad\times
\left(\alpha_{\Omega\{T,\lambda\}}^{(\sigma)*}
\alpha_{\Omega\{T',\lambda'\}}^{(\sigma')}
-\beta_{\Omega\{T,\lambda\}}^{(\sigma)*}
\beta_{\Omega\{T',\lambda'\}}^{(\sigma')}
\right)\dTlambda \cdot\eta_{AA'}\nn\\
&&=\delta_{\sigma,\sigma'}\delta(\Omega-\Omega')
\dTlambda \cdot\eta_{AA'}\nn\\
&&\qquad\qquad\times
\left(\alpha_{\Omega\{T,\lambda\}}^{(\sigma)*}
\alpha_{\Omega\{T,-\lambda\}}^{(\sigma)}
-\beta_{\Omega\{T,\lambda\}}^{(\sigma)*}
\beta_{\Omega\{T,-\lambda\}}^{(\sigma)}
\right)\ ,
\end{eqnarray}
where the $k$-integration has been carried out by the change of the
integration variables from $k$ to $y$ through $k=\mu\sinh y$,
which implies $\omega_k=\mu\cosh y$,
$(\omega_k+k)/(\omega_k-k)=e^{2y}$, and $dk/\omega_k=dy$:
\begin{equation}
\int_{-\infty}^\infty\frac{dk}{2\pi\omega_k}
\left(\frac{\omega_k+k}{\omega_k-k}\right)^{\!
i\left(\sigma\Omega-\sigma'\Omega'\right)/2}
\!\!=\int_{-\infty}^\infty\! \frac{dy}{2\pi}\,
e^{i\left(\sigma\Omega-\sigma'\Omega'\right)y}
=\delta\left(\sigma\Omega-\sigma'\Omega'\right)\ .
\label{eq:k-int}
\end{equation}
Similarly for $(u^*,u)$, we have
\begin{eqnarray}
&&\left(u_{\Omega\{T,\lambda\}A}^{(\sigma)*},
u_{\Omega'\{T',\lambda'\}A'}^{(\sigma')} \right)
=\delta_{\sigma,-\sigma'}\delta(\Omega-\Omega')
\dTlambda \cdot\left(\phi_A^*,\phi_{A'}\right)\nn\\
&&\hspace*{5cm}\times
\left(\beta_{\Omega\{T,\lambda\}}^{(\sigma)}
\alpha_{\Omega\{T,-\lambda\}}^{(-\sigma)}
-\alpha_{\Omega\{T,\lambda\}}^{(\sigma)}
\beta_{\Omega\{T,-\lambda\}}^{(-\sigma)}
\right)\ .
\label{eq:(u*,u)}
\end{eqnarray}
Therefore, the present Rindler modes satisfy the orthonormality,
\begin{eqnarray}
&&\left(u_{\Omega\{T,\lambda\}A}^{(\sigma)},
u_{\Omega'\{T',\lambda'\}A'}^{(\sigma')} \right)
=\delta_{\sigma,\sigma'}\delta(\Omega-\Omega')
\dTlambda \cdot\eta_{AA'}\ ,\nn\\
&&\left(u_{\Omega\{T,\lambda\}A}^{(\sigma)*},
u_{\Omega'\{T',\lambda'\}A'}^{(\sigma')*} \right)
=-(-)^{\abs{A}}\delta_{\sigma,\sigma'}\delta(\Omega-\Omega')
\dTlambda \cdot\eta_{A'A}\ ,
\label{eq:(u,u)}\\
&&\left(u_{\Omega\{T,\lambda\}A}^{(\sigma)},
u_{\Omega'\{T',\lambda'\}A'}^{(\sigma')*} \right)
=
\left(u_{\Omega\{T,\lambda\}A}^{(\sigma)*},
u_{\Omega'\{T',\lambda'\}A'}^{(\sigma')} \right)
=0\ ,\nn
\end{eqnarray}
if the following conditions hold for
$\alpha_{\Omega\{T,\lambda\}}^{(\sigma)}$ and
$\beta_{\Omega\{T,\lambda\}}^{(\sigma)}$:
\begin{eqnarray}
&&\alpha_{\Omega\{T,\lambda\}}^{(\sigma)*}
\alpha_{\Omega\{T,-\lambda\}}^{(\sigma)}
-\beta_{\Omega\{T,\lambda\}}^{(\sigma)*}
\beta_{\Omega\{T,-\lambda\}}^{(\sigma)}=1\ ,
\label{eq:aa-bb=1}\\[5pt]
&&\beta_{\Omega\{T,\lambda\}}^{(\sigma)}
\alpha_{\Omega\{T,-\lambda\}}^{(-\sigma)}
-\alpha_{\Omega\{T,\lambda\}}^{(\sigma)}
\beta_{\Omega\{T,-\lambda\}}^{(-\sigma)}=0\ .
\label{eq:ba-ab=0}
\end{eqnarray}

\subsection{Wedge condition of the Rindler modes}
\label{sec:BC_u-I}

To complete constructing the string Rindler modes, we have to impose
the wedge condition (condition (III)); the vanishing of the modes in
one of the Rindler wedges $\pm$. This is, however, a difficult task
since string is an extended object as we explained in detail
in Sec.\ 1.
In this section we shall fix this problem of imposing the wedge
condition by taking the most naive definition of the wedges.
This is to define the $\pm$ wedges according to whether the Minkowski
CM coordinate $x^\mu$ is in the $+$ wedge or the $-$ one:
\begin{equation}
u_{\Omega\{T,\lambda\}A}^{(\pm)}=0\quad
\mbox{if } \pm x^1 <0\ .
\label{eq:BC_u}
\end{equation}
This is the same wedge condition as in the naive approach
mentioned at the end of Sec.\ 1.

Before imposing the wedge condition on
$u_{\Omega\{T,\lambda\}A}^{(\sigma)}$, we first have to
introduce a regularization factor into (\ref{eq:u_Tlambda})
necessary to make the $k$-integration well-defined.
The same regularization also justifies the discarding of the surface
term which we did in deriving eq.\ (\ref{eq:Eq_for_alpha}) using
integration by parts.
Consider the $k$-integration in (\ref{eq:u_Tlambda}):
\begin{eqnarray}
&&\int_{-\infty}^\infty\! \frac{dk}{\sqrt{2\pi\omega_k}}
\left(\frac{\omega_k+k}{\omega_k-k}\right)^{\!
-\left(i\sigma\Omega+\sum_n\lambda_n\right)/2}\kern-4pt
U_k(x^0,x^1)\nn\\
&&\quad =\frac{1}{2\sqrt{2}\pi}\int^\infty_{-\infty}\!dy\,
e^{-\left(i\sigma\Omega+\sum_n\lambda_n\right)y
+ i\mu\oxi\sinh(y-\oeta)}\nn\\
&&\quad
=\frac{1}{2\sqrt{2}\pi}
e^{-\left(i\sigma\Omega+\sum_n\lambda_n\right)\oeta}
\int^\infty_0\frac{du}{u}
u^{-\left(i\sigma\Omega+\sum_n\lambda_n\right)}
\exp\left\{\frac{i}{2}\mu\oxi\left(u-\frac{1}{u}\right)\right\}\ ,
\label{eq:3int}
\end{eqnarray}
where $(\oxi,\oeta)$ is related to the Minkowski CM coordinate
$(x^0,x^1)$ by
\begin{equation}
x^0=\oxi\sinh\oeta ,\quad
x^1=\oxi\cosh\oeta ,
\label{eq:oeta_oxi_again}
\end{equation}
(this is the same as eq.\ (\ref{eq:oeta_oxi})),
and the three integration variables $k$, $y$ and $u$ are related by
$k=\mu\sinh y$ and $u=e^y$ together with the shift $y\to y+h$.
The integrations of (\ref{eq:3int}) are, however, well-defined only
when $\abs{\Re\left(i\sigma\Omega+\sum_n\lambda_n\right)}<1$, which
implies $\sum_n\lambda_n=0$. In order to make the integration
(\ref{eq:3int}) well-defined for any $\sum_n\lambda_n$, we multiply
the integrand by the regularization factor,
\begin{equation}
\exp\left(-\epsilon\omega_k\right)
=\exp\left(-\epsilon\mu\cosh y\right)
=\exp\left\{-\frac{1}{2}\epsilon\mu\left(u+\frac{1}{u}\right)
\right\}\ ,
\label{eq:regfac}
\end{equation}
and take the limit $\epsilon\to 0$ after the integration.
Then, the integral is reduced to the modified Bessel function
by the formula:
\begin{eqnarray}
&&Q_\nu(x)\equiv \lim_{\epsilon\to +0}
\frac{1}{2}\int_0^\infty \frac{du}{u}u^{-\nu}
\exp\left\{\frac{i}{2}x\left(u-\frac{1}{u}\right)
-\epsilon\left(u + \frac{1}{u}\right)\right\}\nn\\[5pt]
&&\phantom{Q_\nu(x)}
=\cases{
e^{-i\pi\nu/2}K_\nu(x) & $x>0$ \cr
\noalign{\vskip5pt}
e^{i\pi\nu/2}K_\nu(-x) & $x<0$\ ,}
\label{eq:Q}
\end{eqnarray}
where the modified Bessel function $K_\nu(x)$ is defined for $x>0$ and
an arbitrary complex $\nu$ by
\begin{equation}
K_\nu(x)=
\frac{1}{2}\int_0^\infty \frac{du}{u}u^{-\nu}
\exp\left\{-\frac{1}{2}x\left(u+\frac{1}{u}\right)\right\} ,
\quad (x>0,\ {}^\forall\nu\in\bm{C})\ .
\label{eq:defK}
\end{equation}
(See Chapter 6 of ref.\ \cite{Watson} for the details.)
Owing to the regularization factor, the $u$-integration along the
positive real axis in (\ref{eq:Q}) can be deformed to the integration
along the imaginary axis; $u\to i\sgn(x) u$. Without the
regularization, this contour deformation is allowed only
when $\abs{\Re\nu}<1$ due to the singularities at $\abs{u}\to 0$ and
$\infty$.
Note also that $Q_\nu(x)$ enjoys the property,
\begin{equation}
Q_{-\nu}(x)=Q_\nu(-x)\ ,
\label{eq:Q_-nu}
\end{equation}
as is seen by the change of the integration variables $u\to 1/u$.

Having finished mathematical preparation,
the Rindler wave function (\ref{eq:u_Tlambda}) with the regularization
is rewritten using (\ref{eq:Q}) as
\begin{eqnarray}
&&u_{\Omega\{T,\lambda\}A}^{(\sigma)}=
\frac{e^{-\Lambda\oeta}}{\sqrt{2}\,\pi}
\left(\alpha_{\Omega\{T,\lambda\}}^{(\sigma)}
Q_\Lambda\left(\mu\oxi\right)
+ \beta_{\Omega\{T,\lambda\}}^{(\sigma)}
Q_\Lambda\left(-\mu\oxi\right)
\right)\Phi_{\{T,\lambda\}}\cdot\phi_A \nn\\[5pt]
&&\phantom{u_{\Omega\{T,\lambda\}A}^{(\sigma)}}
\propto
\left(\alpha_{\Omega\{T,\lambda\}}^{(\sigma)}
e^{-i\pi\Lambda\sgn(\oxi)/2}
+
\beta_{\Omega\{T,\lambda\}}^{(\sigma)}
e^{i\pi\Lambda\sgn(\oxi)/2}\right)
K_{\Lambda}\!\left(\mu|\oxi|\right)\ ,
\label{eq:u-propto}
\end{eqnarray}
with
\begin{equation}
\Lambda\equiv i\sigma\Omega+\sum_n\lambda_n\ .
\label{eq:Delta}
\end{equation}
Now we impose the wedge condition defined by the sign of $\oxi$:
$u_{\Omega\{T,\lambda\}A}^{(+)}$
($u_{\Omega\{T,\lambda\}A}^{(-)}$) should vanish when $\oxi<0$
($\oxi>0$).
Eq.\ (\ref{eq:u-propto}) implies that this condition is satisfied if
$\alpha_{\Omega\{T,\lambda\}}^{(\sigma)}$ and
$\beta_{\Omega\{T,\lambda\}}^{(\sigma)}$ are related by
\begin{equation}
\alpha_{\Omega\{T,\lambda\}}^{(\sigma)}
e^{i\pi\sigma\Lambda/2}
+ \beta_{\Omega\{T,\lambda\}}^{(\sigma)}
e^{-i\pi\sigma\Lambda/2}=0\ ,
\label{eq:alpha+beta=0}
\end{equation}
and hence
\begin{equation}
\beta_{\Omega\{T,\lambda\}}^{(\sigma)}
=-(-)^{\sum_n\lambda_n}e^{-\pi\Omega}
\alpha_{\Omega\{T,\lambda\}}^{(\sigma)}\ .
\label{eq:beta=alpha}
\end{equation}
Eq.\ (\ref{eq:beta=alpha}) together with the normalization condition
(\ref{eq:aa-bb=1}) determines
$\alpha_{\Omega\{T,\lambda\}}^{(\sigma)}$
and $\beta_{\Omega\{T,\lambda\}}^{(\sigma)}$ as
\begin{equation}
\alpha_{\Omega\{T,\lambda\}}^{(\sigma)}
=\sqrt{N(\Omega)+1}\ ,
\quad
\beta_{\Omega\{T,\lambda\}}^{(\sigma)}
=-(-)^{\sum_n\lambda_n}\sqrt{N(\Omega)}\ ,
\label{eq:alphabeta_final}
\end{equation}
where $N(\Omega)$ is the Bose distribution function at temperature
$T=1/2\pi{}k_{\rm B}$ ($k_{\rm B}$ is the Boltzmann's constant):
\begin{equation}
N(\Omega)\equiv\left(e^{2\pi\Omega}-1\right)^{-1}\ .
\label{eq:N(Omega)}
\end{equation}
The other condition (\ref{eq:ba-ab=0}) is automatically satisfied by
eq.\ (\ref{eq:alphabeta_final}).
$\alpha_{\Omega\{T,\lambda\}}^{(\sigma)}$ and
$\beta_{\Omega\{T,\lambda\}}^{(\sigma)}$ as given by
(\ref{eq:alphabeta_final}) are not the unique solution of
(\ref{eq:aa-bb=1}): there is an arbitrariness of multiplying them
by a factor $C(\{\lambda\})$ with the property
$C^*(\{\lambda\})C(\{-\lambda\})=1$. However, this arbitrariness does
not have any physical meaning since the whole wave function
$u_{\Omega\{T,\lambda\}A}^{(\sigma)}$ is multiplied by
$C(\{\lambda\})$, and hence can be absorbed into the definition of the
creation/annihilation operators in the expansion of the string field
$\Phi$ in terms of the Rindler modes (c.f., eq.\ (\ref{eq:Phi=ub})).
Substituting eq.\ (\ref{eq:alphabeta_final}) into (\ref{eq:u_Tlambda}),
our string Rindler mode is finally given by
\begin{eqnarray}
&&u_{\Omega\{T,\lambda\}A}^{(\sigma)}=
\int_{-\infty}^\infty\! \frac{dk}{\sqrt{2\pi\omega_k}}
e^{-\epsilon\omega_k}
\left(\frac{\omega_k+k}{\omega_k-k}\right)^{\!
-\left(i\sigma\Omega+\sum_n\lambda_n\right)/2}
\nn\\
&&\hspace*{2.3cm}\times
\left(\sqrt{N(\Omega)+1}\,U_k
-(-)^{\sum_n\lambda_n}\sqrt{N(\Omega)}\,U_k^*\right)
\Phi_{\{T,\lambda\}}
\cdot\phi_A\ .
\label{eq:u_Tlambda_final}
\end{eqnarray}

\subsection{Comparison with the Rindler mode in the naive approach}

We finish this section by comparing our Rindler mode
$u_{\Omega\{T,\lambda\}A}^{(\sigma)}$ (\ref{eq:u_Tlambda_final})
with another Rindler mode in the naive approach which
treats $\oeta$ of (\ref{eq:oeta_oxi_again}) as the Rindler time (see
Sec.\ 1).
Let us denote the latter mode by $v_{\Omega\{M,N\}A}^{(\sigma)}$.
Corresponding to eq.\ (\ref{eq:u_Tlambda_final}) for
$u_{\Omega\{T,\lambda\}A}^{(\sigma)}$,
$v_{\Omega\{M,N\}A}^{(\sigma)}$ is given in terms of the components
of the string Minkowski mode as
\begin{eqnarray}
&&v_{\Omega\{M,N\}A}^{(\sigma)}=
\int_{-\infty}^\infty\! \frac{dk}{\sqrt{2\pi\omega_k}}
\left(\frac{\omega_k+k}{\omega_k-k}\right)^{\!
-i\sigma\Omega/2}
\nn\\
&&\hspace*{3cm}\times
\left(\sqrt{N(\Omega)+1}\,U_k
-\sqrt{N(\Omega)}\,U_k^*\right)
\Psi_{\{M,N\}}
\cdot\phi_A\ .
\label{eq:v}
\end{eqnarray}
Besides the requirement that $v_{\Omega\{M,N\}A}^{(\sigma)}$
satisfy the the string wave equation
$L v_{\Omega\{M,N\}A}^{(\sigma)}=0$,
$v_{\Omega\{M,N\}A}^{(\sigma)}$ has been constructed by
imposing the following three conditions.
First it carries the Rindler energy $\Omega$ with respect to the
$\oeta$ time:
\begin{equation}
\henbi{}{\oeta}v_{\Omega\{M,N\}A}^{(\sigma)}=-i\sigma\Omega
v_{\Omega\{M,N\}A}^{(\sigma)}\ .
\label{eq:dv/deta_M}
\end{equation}
Second, it satisfies the normalization condition:
\begin{eqnarray}
&&\left(v_{\Omega\{M,N\}A}^{(\sigma)},
v_{\Omega'\{M',N'\}A'}^{(\sigma')}\right)
=\delta_{\sigma,\sigma'}
\delta\left(\Omega-\Omega'\right)\delta_{\{M,N\},\{M',N'\}}
(-)^{\sum_n{M_n}}\eta_{AA'}\ ,\nn\\
&&\left(v_{\Omega\{M,N\}A}^{(\sigma)*},
v_{\Omega'\{M',N'\}A'}^{(\sigma')*}\right)
=-(-)^{|A|}\delta_{\sigma,\sigma'}\delta\left(\Omega-\Omega'\right)
\delta_{\{M,N\},\{M',N'\}}(-)^{\sum_n{M_n}}\eta_{A'A}\ ,\nn\\
&&\left(v_{\Omega\{M,N\}A}^{(\sigma)},
v_{\Omega'\{M',N'\}A'}^{(\sigma')*}\right)=0\ .
\label{eq:(v,v)}
\end{eqnarray}
Finally, $v_{\Omega\{M,N\}A}^{(\sigma)}$ is subject to the wedge
condition specified by the Minkowski CM coordinate $x^\mu$:
\begin{equation}
v_{\Omega\{M,N\}A}^{(\pm)}=0\quad \mbox{if }\pm x^1 <0\ .
\label{eq:BC_v}
\end{equation}
The wedge condition (\ref{eq:BC_v}) is the same as (\ref{eq:BC_u})
which we imposed on $u_{\Omega\{T,\lambda\}A}^{(\sigma)}$ in the
last subsection.

Then, let us consider the inner product between
$u_{\Omega\{T,\lambda\}A}^{(\sigma)}$ and the present
$v_{\Omega\{M,N\}A}^{(\sigma)}$. However, this inner product is
ill-defined:
\begin{eqnarray}
&&\left(v_{\Omega'\{M,N\}A'}^{(\sigma')},
u_{\Omega\{T,\lambda\}A}^{(\sigma)}\right)
=\int_{-\infty}^\infty\frac{dk}{2\pi\omega_k}
\left(\frac{\omega_k +k}{\omega_k -k}\right)^{
i\left(\sigma'\Omega'-\sigma\Omega\right)/2 -\sum_n\lambda_n/2}\nn\\
&&\qquad\quad\times
\left(\sqrt{\left(N(\Omega')+1\right)\left(N(\Omega)+1\right)}
-(-)^{\sum_n\lambda_n}\sqrt{N(\Omega')N(\Omega)}
\right)\nn\\
&&\qquad\quad\times
\prod_n\alpha_{T_n,-\lambda_n}(M_n)\delta_{M_n+N_n,T_n}
\cdot\eta_{A'A}\ ,
\end{eqnarray}
which is divergent at either $k=\infty$ or $-\infty$ unless
$\sum_n\lambda_n=0$.
Therefore, we conclude that the Rindler quantization of SFT using
$u_{\Omega\{T,\lambda\}A}^{(\sigma)}$ and the one using
$v_{\Omega\{M,N\}A}^{(\sigma)}$ are not equivalent.

\section{Rindler quantization of SFT and string thermalization}

In the previous section we have constructed the string Rindler mode
$u_{\Omega\{T,\lambda\}A}^{(\sigma)}$,
eq.\ (\ref{eq:u_Tlambda_final}).
Though this Rindler mode is not a fully satisfactory one in the point
that the wedge condition is specified by the Minkowski CM
coordinate, we shall in this section
carry out the Rindler quantization of free
SFT using this Rindler mode and then study the string thermalization.

\subsection{Rindler quantization of free SFT}

The Rindler quantization starts with defining the
canonical momentum $\Pi_{\rm R}$ conjugate to $\Phi$ and then imposing
the canonical commutation relation. The action of free SFT expressed
in terms of the Rindler string coordinate reads
(c.f., eqs.\ (\ref{eq:S_by_Y}), (\ref{eq:G_munu_R}) and (\ref{eq:G_R})
in Appendix \ref{app:innprod_R}),
\begin{eqnarray}
&&S=\frac{\pi}{2}
\int\calD\xi(\sigma)\calD\eta(\sigma)\calD X_\bot(\sigma)
\calD(\mbox{ghosts})
\prod_{\sigma'}\abs{\xi(\sigma')}\nn\\
&&\qquad\times
\int_0^\pi d\sigma\Biggl\{
\frac{1}{\xi^2(\sigma)}\left(\hanbi{\Phi}{\eta(\sigma)}\right)^2
-\left(\hanbi{\Phi}{\xi(\sigma)}\right)^2
+\left[
\left(\xi(\sigma)\right)^2\left(\eta'(\sigma)\right)^2
- \left(\xi'(\sigma)\right)^2\right]\Phi^2
\nn\\
&&\hspace*{4cm}
+(\mbox{transverse and ghost coordinates part})\Biggr\}\ .
\label{eq:S_R}
\end{eqnarray}
Taking $\eta_0$ of eq.\ (\ref{eq:eta_0}) as the time variable for
quantization, the canonical momentum $\Pi_{\rm R}$ is given by
\begin{equation}
\Pi_{\rm R}
\equiv \hanbi{S}{(\p\Phi/\p \eta_0)}
=\prod_{\sigma'}\abs{\xi(\sigma')}\int_0^\pi\! d\sigma
\frac{1}{\xi(\sigma)^2}
\hanbi{\Phi}{\eta(\sigma)}\ .
\label{eq:Pi_R}
\end{equation}
Then we impose the equal-$\eta_0$ commutation relation
\begin{eqnarray}
&&\left.\left[\Phi[\eta,\xi,X_\bot,c,\ac],
\Pi_R[\til\eta,\til\xi,\til{X}_\bot,\til{c},\til{\ac}]\right]
\right\vert_{\eta_0=\til\eta_0}
=\nn\\
&&\hspace*{2.5cm}
i\prod_{\sigma}\widehat{\delta}\left(\eta(\sigma)-\til\eta(\sigma)\right)
\delta\left(\xi(\sigma)-\til\xi(\sigma)\right)
\delta\left(X_\bot,c,\ac\ \mbox{-part}\right) ,
\label{eq:CCR_R}
\end{eqnarray}
where the hat on
$\widehat{\delta}\left(\eta(\sigma)-\til\eta(\sigma)\right)$
implies that the delta function for the time variable,
$\delta(\eta_0-\til\eta_0)$, is omitted.

Corresponding to eq.\ (\ref{eq:Phi=Ua}) in the Minkowski quantization,
let us expand the string field $\Phi$ in terms of the string Rindler
modes $u_{\Omega\{T,\lambda\}A}^{(\sigma)}$ and its
complex conjugate:
\begin{equation}
\Phi=\sum_{\sigma=\pm}
\int_0^\infty d\Omega \sum_{\{T,\lambda\}}\sum_A
\left(u_{\Omega\{T,\lambda\}A}^{(\sigma)}
b_{\Omega\{T,\lambda\}}^{(\sigma)A}
+ b_{\Omega\{T,\lambda\}}^{(\sigma)A\dagger}
u_{\Omega\{T,\lambda\}A}^{(\sigma)*}
\right)\ .
\label{eq:Phi=ub}
\end{equation}
The orthonormality of the Rindler modes (\ref{eq:(u,u)}) allows us to
express $b$ and $b^\dagger$ in terms of $\Phi$:
\begin{eqnarray}
&&b_{\Omega\{T,\lambda\}}^{(\sigma)A}
=\sum_{B}\eta^{AB}
\left(u_{\Omega\{T,-\lambda\}B}^{(\sigma)}\, , \Phi\right)\ ,\nn\\
&&b_{\Omega\{T,\lambda\}}^{(\sigma)A\dagger}
=-\sum_{B}
\left(u_{\Omega\{T,-\lambda\}B}^{(\sigma)*}\, ,
\Phi\right)\eta^{BA}\ .
\label{eq:b=(u,Phi)}
\end{eqnarray}
The (anti-)commutation relation between $b$ and $b^\dagger$ is
obtained by identifying the inner product in eq.\
(\ref{eq:b=(u,Phi)}) as the Rindler one (\ref{eq:innprod_R}), and
using the commutation relation (\ref{eq:CCR_R}). Then we find
\begin{eqnarray}
\CR{b_{\Omega\{T,\lambda\}}^{(\sigma)A},
b_{\Omega'\{T',\lambda'\}}^{(\sigma')A'\dagger}}
\!&=&\!\sum_{B,B'}\eta^{AB}
\left(u_{\Omega\{T,\lambda\}B}^{(\sigma)},
u_{\Omega'\{T',\lambda'\}B'}^{(\sigma')} \right)\eta^{B'A'}\nn\\
\!&=&\!\delta_{\sigma,\sigma'}\delta(\Omega-\Omega')
\dTlambda \cdot\eta^{AA'}\ ,
\label{eq:[b,b+]}
\end{eqnarray}
and
\begin{equation}
\CR{b_{\Omega\{T,\lambda\}}^{(\sigma)A},
b_{\Omega'\{T',\lambda'\}}^{(\sigma')A'}}
=
\CR{b_{\Omega\{T,\lambda\}}^{(\sigma)A\dagger},
b_{\Omega'\{T',\lambda'\}}^{(\sigma')A'\dagger}}
=0\ .
\label{eq:[b,b]}
\end{equation}

The Rindler Hamiltonian which is the generator of the $\eta_0$
translation and hence satisfies
\begin{equation}
\left[H_{\rm R}, \Phi\right]=-i\henbi{}{\eta_0}\Phi\ ,
\label{eq:[H_R,Phi]}
\end{equation}
is given as
\begin{equation}
H_{\rm R}=H_{\rm R}^{(+)} - H_{\rm R}^{(-)}\ ,
\label{eq:H=H-H}
\end{equation}
where $H_{\rm R}^{(\pm)}$ is the Hamiltonian in the $\pm$-wedges:
\begin{equation}
H_{\rm R}^{(\sigma)}=
\int_0^\infty\! d\Omega\,\Omega\!
\sum_{\{T,\lambda\}}\sum_{A,B}
b_{\Omega\{T,\lambda\}}^{(\sigma)A\dagger}\eta_{AB}
b_{\Omega\{T,-\lambda\}}^{(\sigma)B}\ .
\label{eq:H_R(pm)}
\end{equation}
The Rindler vacuum state $\rket{0}$ is defined as the state annihilated
by both $b^{(\pm)}$:
\begin{equation}
b_{\Omega\{T,\lambda\}}^{(\pm)A}\rket{0}=0\ .
\label{eq:vac_R}
\end{equation}
This is the groundstate of the Hamiltonians $H_{\rm R}^{(\pm)}$.
In Appendix \ref{app:QBket0_R=0}, we show that the Rindler vacuum
defined by eq.\ (\ref{eq:vac_R}) is a BRST invariant physical state.

\subsection{Bogoliubov coefficients and string thermalization}

Our task in this subsection is first to obtain the relationship
between the Minkowski creation/annihilation operators $\{a,a^\dagger\}$
and the Rindler ones $\{b,b^\dagger\}$.
The relationship between these two sets of creation/annihilation
operators is obtained by equating the two expressions of the
string field $\Phi$, eqs.\ (\ref{eq:Phi=Ua}) and (\ref{eq:Phi=ub}),
and using the orthonormality of the string modes, either eqs.\
(\ref{eq:a=(U,Phi)}) or (\ref{eq:b=(u,Phi)}).
Therefore, the Rindler annihilation operator expressed in terms of the
Minkowski set $\{a,a^\dagger\}$ is given as
\begin{eqnarray}
&&b_{\Omega\{T,\lambda\}}^{(\sigma)A}=
\sum_{{\scriptstyle \{M,N\}\atop\scriptstyle M+N=T}}
\int_{-\infty}^\infty\! \frac{dk}{\sqrt{2\pi\omega_k}}
\left(\frac{\omega_k+k}{\omega_k-k}\right)^{\!
\left(i\sigma\Omega+\sum_n\lambda_n\right)/2}
\prod_n\alpha_{T_n,\lambda_n}(M_n)
\nn\\
&&\hspace*{1.5cm}\times
\left(\sqrt{N(\Omega)+1}\,a_{k\{M,N\}}^A
+(-)^{\sum_n\lambda_n}\sqrt{N(\Omega)}
\sum_{B,C}\eta^{AB}(\phi_{B},\phi_C^*)(-)^{|C|}\,
a_{k\{M,N\}}^{C\dagger} \right) ,\nn\\
\label{eq:b=a+a^+}
\end{eqnarray}
where we have used the formulas
\begin{eqnarray}
&&\left(u_{\Omega\{T,\lambda\}A}^{(\sigma)},
U_{k\{M,N\}B}\right)
=\sqrt{\frac{N(\Omega)+1}{2\pi\omega_k}}
\left(\frac{\omega_k+k}{\omega_k-k}\right)^{\!
\left(i\sigma\Omega-\sum_n\lambda_n\right)/2}
\prod_n\alpha_{T_n,-\lambda_n}(M_n)\cdot \eta_{AB}\ ,\nn\\ \\
&&\left(u_{\Omega\{T,\lambda\}A}^{(\sigma)},
U_{k\{M,N\}B}^*\right)
=(-)^{\sum_n\lambda_n}
\sqrt{\frac{N(\Omega)}{2\pi\omega_k}}
\left(\frac{\omega_k+k}{\omega_k-k}\right)^{\!
\left(i\sigma\Omega-\sum_n\lambda_n\right)/2}\nn\\
&&\hspace*{6cm}
\times\prod_n\alpha_{T_n,-\lambda_n}(M_n)\cdot
\left(\phi_A,\phi_B^*\right)\ ,
\end{eqnarray}
obtainable from eqs.\ (\ref{eq:u_Tlambda_final}),
(\ref{eq:Phi_Tlambda}), (\ref{eq:(Psi_MN,Psi_MN)}) and
(\ref{eq:alpha_T(-lambda)}).
As in particle field theory \cite{BD,bible}, it is convenient
to define another set of Minkowski annihilation operators
$d_{\Omega\{T,\lambda\}}^{(\sigma)A}$,
\begin{eqnarray}
d_{\Omega\{T,\lambda\}}^{(\sigma)A}=
\sum_{{\scriptstyle \{M,N\}\atop\scriptstyle M+N=T}}
\int_{-\infty}^\infty\! \frac{dk}{\sqrt{2\pi\omega_k}}
\left(\frac{\omega_k+k}{\omega_k-k}\right)^{\!
\left(i\sigma\Omega+\sum_n\lambda_n\right)/2}\!\!
\prod_n\alpha_{T_n,\lambda_n}(M_n)\cdot a_{k\{M,N\}}^A\ ,
\end{eqnarray}
which annihilate the Minkowski vacuum,
\begin{equation}
d_{\Omega\{T,\lambda\}}^{(\sigma)A}\mket{0}=0\ ,
\label{eq:dvac_M=0}
\end{equation}
and satisfy the (anti-)commutation relations,
\begin{eqnarray}
&&\CR{d_{\Omega\{T,\lambda\}}^{(\sigma)A},
d_{\Omega'\{T',\lambda'\}}^{(\sigma')A'\dagger}}
=\delta_{\sigma,\sigma'}\delta(\Omega-\Omega')
\dTlambda \cdot\eta^{AA'}\ ,\nn\\
&&\CR{d_{\Omega\{T,\lambda\}}^{(\sigma)A},
d_{\Omega'\{T',\lambda'\}}^{(\sigma')A'}}=
\CR{d_{\Omega\{T,\lambda\}}^{(\sigma)A\dagger},
d_{\Omega'\{T',\lambda'\}}^{(\sigma')A'\dagger}}=0\ .
\end{eqnarray}
Then, the Rindler operators $\{b,b^\dagger\}$ expressed in terms of new
Minkowski ones $\{d,d^\dagger\}$ is given by
\begin{eqnarray}
&&b_{\Omega\{T,\lambda\}}^{(\sigma)A}=
\sqrt{N(\Omega)+1}\,d_{\Omega\{T,\lambda\}}^{(\sigma)A}
+(-)^{\sum_n\lambda_n}\sqrt{N(\Omega)}
\sum_{B,C}\eta^{AB}(\phi_{B},\phi_C^*)(-)^{|C|}\,
d_{\Omega\{T,\lambda\}}^{(-\sigma)C\dagger} ,\nn\\
\label{eq:b=d+d^+}\\
&&b_{\Omega\{T,\lambda\}}^{(\sigma)A\dagger}=
\sqrt{N(\Omega)+1}\,d_{\Omega\{T,\lambda\}}^{(\sigma)A\dagger}
+(-)^{\sum_n\lambda_n}\sqrt{N(\Omega)}
\sum_{B,C}d_{\Omega\{T,\lambda\}}^{(-\sigma)C}
(-)^{|C|}(\phi_C^*,\phi_B)\eta^{BA} .\nn
\end{eqnarray}
Therefore, the Rindler vacuum $\rket{0}$ defined by (\ref{eq:vac_R})
is formally expressed in terms of the Minkowski vacuum $\mket{0}$ and
$d_{\Omega\{T,\lambda\}}^{(\pm)A\dagger}$ as
\begin{equation}
\rket{0}\propto\exp\left(
-\int_0^\infty d\Omega\sum_{\{T,\lambda\}}\sum_{A,B}
e^{-\pi\Omega}(-)^{\sum_n\lambda_n}
d_{\Omega\{T,\lambda\}}^{(+)A\dagger}
(\phi_A,\phi_B^*)(-)^{|B|}
d_{\Omega\{T,-\lambda\}}^{(-)B\dagger}
\right)\mket{0}\ ,
\label{eq:vac_R=exp_vac_M}
\end{equation}
up to the normalization.

As an application of the Bogoliubov transformation relation
(\ref{eq:b=d+d^+}), let us consider the Minkowski vacuum expectation
value of the Rindler Hamiltonian $H_{\rm R}^{(+)}$ in the $+$ wedge,
namely, $\mbra{0}H_{\rm R}^{(+)}\mket{0}$.
Only the last terms of eq.\ (\ref{eq:b=d+d^+}) contribute to this
expectation value, and we get
\begin{equation}
\mbra{0}H_{\rm R}^{(+)}\mket{0}
=\sum_{\{T,\lambda\}}\sum_A(-)^{|A|}\delta^A_A\cdot
\delta(\Omega-\Omega)\int_0^\infty\! d\Omega\,\Omega N(\Omega)\ ,
\label{eq:vac_MH_Rvac_M}
\end{equation}
where in the course of the calculation we have used the formulas
\begin{eqnarray}
&&\phi=\sum_{A,B}\phi_A\eta^{AB}(\phi_B,\phi)\quad
\mbox{for } {}^\forall\phi\ ,
\label{eq:complt_phi}\\
&&\phi=\sum_{A,B}\phi_A^*(-)^{|A|}
\eta^{BA}(\phi_B^*,\phi)\quad\mbox{for } {}^\forall\phi\ ,
\label{eq:complt_phi*}\\
&&(\phi_A^*,\phi_B^*)=(-)^{|A|}\eta_{BA}\ .
\end{eqnarray}
Eqs.\ (\ref{eq:complt_phi}) and (\ref{eq:complt_phi*}) are the
completeness relations of the transverse wave functions
$\{\phi_A\}$ and $\{\phi_A^*\}$.

The last part
$\delta(\Omega=0)\int_0^\infty d\Omega\Omega N(\Omega)$
of eq.\ (\ref{eq:vac_MH_Rvac_M}) is the same as
$\mbra{0}H_{\rm R}^{(+)}\mket{0}$ calculated in the free scalar
particle field theory \cite{BD,bible}, and shows the standard Bose
distribution at temperature $T=1/2\pi{}k_{\rm B}$.
The interpretation of the other part on the RHS
of eq.\  (\ref{eq:vac_MH_Rvac_M}) is as follows.
First, since the index $A$ specifies the transverse momentum
$\bm{k}_\bot$ as well as the discrete level numbers, $\delta^A_A$ for a
fixed $A$ is given as
$\delta^A_A=\delta^{24}(\bm{k}_\bot-\bm{k}_\bot)=V_\bot/(2\pi)^{24}$
with $V_\bot\equiv\int d^{24}x$ being the transverse volume.
Then the summation of the sign factor $(-)^{|A|}$ over
$\{T,\lambda\}$ and the transverse index $A$ (except $\bm{k}_\bot$)
counts the number of physical component fields contained in the free
string field $\Phi$ in the Minkowski quantization.
Therefore, we have
\begin{equation}
\sum_{\{T,\lambda\}}\sum_A(-)^{|A|}\delta^A_A
=\mbox{(\# physical component fields)}\times
V_\bot\int \frac{d^{24}k_\bot}{(2\pi)^{24}}\ ,
\label{eq:sum_delta^A_A}
\end{equation}
and hence (\ref{eq:vac_MH_Rvac_M}) is equal to
$\mbra{0}H_{\rm R}^{(+)}\mket{0}$ in the free scalar field theory
multiplied by the number of the physical component fields
in $\Phi$.
If we consider the same expectation value
$\mbra{0}H_{\rm R}^{(+)}\mket{0}$
in the naive approach mentioned at the end of Sec.\ 1, we get the same
result (\ref{eq:vac_MH_Rvac_M}) with $\sum_{\{T,\lambda\}}$ replaced
by $\sum_{\{M,N\}}$.

The (ultraviolet) divergent factor $\delta(\Omega-\Omega)$
in eq.\ (\ref{eq:vac_MH_Rvac_M}), which also appears in particle field
theory, is due to the continuum nature of the Rindler energy $\Omega$.
To regularize this factor in the present SFT case, we need to
introduce something like the horizon regularization of
ref.\ \cite{VegaSanchez,SusskindUglum}.
The physical difference between the Rindler quantization of this paper
and the naive approach may be exposed by such detailed analysis.
This is the subject of our future investigation.

\section{Rindler modes with different wedge conditions}
\label{sec:otherBC}

The wedge condition (\ref{eq:BC_u}) of the Rindler mode
$u_{\Omega\{T,\lambda\}A}^{(\sigma)}$ (\ref{eq:u_Tlambda_final})
which we constructed in Sec.\ \ref{sec:BC_u-I} was specified by the
Minkowski CM coordinate $x^\mu$.
In this section we present an approach to the construction of the
string Rindler mode which is subject to different wedge
conditions. Although our consideration here is still incomplete,
we hope that the analysis in this section will give a hint for a
thorough understanding of the Rindler quantization of SFT.

The Rindler mode we shall construct here is given in the form of eq.\
(\ref{eq:u_Z}), namely, by summing over the $\{T,\lambda\}$ quantum
numbers with a suitable weight.
Using the function $Q_\nu(x)$ defined by eq.\ (\ref{eq:Q}),
$u_{\Omega,Z,A}^{(\sigma)}$ (\ref{eq:u_Z}) is expressed as
\begin{equation}
u_{\Omega,Z,A}^{(\sigma)}=
\frac{1}{\sqrt{2}\,\pi}
\sum_{\{T,\lambda\}}e^{-\Lambda\oeta}
\left(\alpha_{\Omega,Z,\{T,\lambda\}}^{(\sigma)}
Q_\Lambda\!\left(\mu\oxi\right)
+ \beta_{\Omega,Z,\{T,\lambda\}}^{(\sigma)}
Q_{-\Lambda}\!\left(\mu\oxi\right)
\right)\Phi_{\{T,\lambda\}}\cdot\phi_A\ ,
\label{eq:u_Z-1}
\end{equation}
with $\Lambda$ of eq.\ (\ref{eq:Delta}).
We shall carry out the summation over $\{T,\lambda\}$ in
(\ref{eq:u_Z-1}) by assuming a certain $\{T,\lambda\}$ dependence of
the coefficients $\alpha_{\Omega,Z,\{T,\lambda\}}^{(\sigma)}$
and $\beta_{\Omega,Z,\{T,\lambda\}}^{(\sigma)}$.
For this purpose we need a number of mathematical preliminaries.
The first is the following expansion formula
for $Q_\nu(x)$:\footnote{
This formula can be understood from the same expansion formula for the
Bessel functions (see Chapter 5.22 of ref.\ \cite{Watson})
and eq.\ (\ref{eq:Q}) which relates $Q_\nu(x)$ to $K_\nu(\pm x)$.
Although $a$ is restricted to lie in the range $|a|<1$, we employ
eq.\ (\ref{eq:expand_Q}) beyond this restriction since the purpose of
this section is merely to give a hint for a more complete analysis of
the wedge condition of the string Rindler modes.
}
\begin{eqnarray}
&&Q_\nu\left(\sqrt{1+a}\,x\right)
=(1+a)^{\nu/2}\sum_{\ell=0}^\infty\frac{(-)^\ell}{\ell!}
\left(\frac{iax}{2}\right)^\ell Q_{\nu+\ell}(x)\ ,\nn\\
&&\phantom{Q_\nu\left(\sqrt{1+a}\,x\right)}
=(1+a)^{-\nu/2}\sum_{\ell=0}^\infty\frac{1}{\ell!}
\left(\frac{iax}{2}\right)^\ell Q_{\nu - \ell}(x)\ ,
\label{eq:expand_Q}
\end{eqnarray}
for a real $a$. Expressing $\mu\oxi$ as
\begin{equation}
\mu\oxi=\left(1 + \frac{2\pi}{\mt^2}\sum_n nT_n\right)^{1/2}\mt\oxi\ ,
\end{equation}
with $\mt\equiv\sqrt{\bm{k}_\bot^2 + m_A^2}$,
we can apply eq.\ (\ref{eq:expand_Q}) to extract the $\{T\}$
dependence of $Q_{\pm\Lambda}(\mu\oxi)$:
\begin{equation}
\pmatrix{
Q_{\Lambda}\left(\mu\oxi\right)\cr
Q_{-\Lambda}\left(\mu\oxi\right)}=
\left(1 + \frac{2\pi}{\mt^2}\sum_n nT_n\right)^{-\Lambda/2}
\sum_{\ell=0}^\infty\frac{1}{\ell!}
\left(\frac{i\pi}{\mt}\sum_n nT_n\,\oxi\right)^\ell
\pmatrix{
Q_{\Lambda-\ell}\left(\mt\oxi\right)\cr
(-)^\ell Q_{-\Lambda+\ell}\left(\mt\oxi\right)
} .
\label{eq:expand_Q_Delta}
\end{equation}

The second formula we need is the expression of
$\Phi_{\{T,\lambda\}}=\prod_n\Phi_{T_n,\lambda_n}$
in terms of the coordinate $(\oxi_n,\oeta_n)$ defined by
\begin{equation}
x^{\mu=0}_n=\oxi_n\sinh\oeta_n\ ,
\quad
x^{\mu=1}_n=\oxi_n\cosh\oeta_n\ .
\label{eq:oxi_n-oeta_n}
\end{equation}
The desired expression of $\Phi_{T_n,\lambda_n}$ is given by
eq.\ (\ref{eq:Phi(oxi_n,oeta_n)}) in Appendix \ref{app:Two_other} using
the Laguerre polynomial.
For our present purpose, observe that
\begin{eqnarray}
&&\sqrt{
\frac{\left(\frac{\ds T-\abs{\lambda}}{\ds 2}\right)!}
{\left(\frac{\ds T+\abs{\lambda}}{\ds 2}\right)!}}
\,y^{|\lambda|}
L_{\frac{1}{2}\left(T-\abs{\lambda}\right)}^{(\abs{\lambda})}(y^2)
=\sqrt{
\left(\frac{T+\lambda}{2}\right)!
\left(\frac{T-\lambda}{2}\right)!}\nn\\
&&\hspace*{12ex}\times
\sum_{p=0}^{N(T,\lambda)}(-)^{N(T,\lambda)-p}
\left[p!
\left(\frac{T+\lambda}{2}-p\right)!
\left(\frac{T-\lambda}{2}-p\right)!
\right]^{-1} y^{T-2p}\ ,
\label{eq:L(y^2)}
\end{eqnarray}
where $N(T,\lambda)$ is defined by
\begin{equation}
N(T,\lambda)=\Half\left(T-\abs{\lambda}\right)\ ,
\label{eq:N(T,lambda)}
\end{equation}
and the summation variable $p$ is related to $r$ in eq.\ (\ref{eq:Laguerre})
by $p=N(T,\lambda)-r$.

Taking into account eqs.\ (\ref{eq:expand_Q_Delta}) and
(\ref{eq:L(y^2)}), let us assume the following form for the
coefficients
$\alpha_{\Omega,Z,\{T,\lambda\}}^{(\sigma)}$ and
$\beta_{\Omega,Z,\{T,-\lambda\}}^{(\sigma)}$ in eq.\ (\ref{eq:u_Z-1}):
\begin{eqnarray}
&&\pmatrix{\alpha_{\Omega,Z,\{T,\lambda\}}^{(\sigma)}\cr
\beta_{\Omega,Z,\{T,-\lambda\}}^{(\sigma)}
}=\sqrt{2}\pi
\left(1 + \frac{2\pi}{\mt^2}\sum_n nT_n\right)^{-\Lambda/2}\nn\\
&&\qquad\qquad\times
\prod_n (-)^{N(T_n,\lambda_n)}\left[\frac{n}{2}
\left(\frac{T_n+\lambda_n}{2}\right)!
\left(\frac{T_n-\lambda_n}{2}\right)!\right]^{-1/2}
\!C_n^{T_n} D_n^{\lambda_n}
\pmatrix{\alpha_{\Omega}^{(\sigma)}\cr \beta_{\Omega}^{(\sigma)}}
\label{eq:assume_alpha&beta}\ ,
\end{eqnarray}
where $C_n$, $D_n$, $\alpha_\Omega^{(\sigma)}$ and
$\beta_\Omega^{(\sigma)}$ will be determined later.
Note the minus sign of the index $\lambda$ for
$\beta_{\Omega,Z,\{T,-\lambda\}}^{(\sigma)}$ on the LHS.
Most of the factors on the RHS of eq.\ (\ref{eq:assume_alpha&beta})
are introduced to cancel the front factors on the RHSs of
eqs.\ (\ref{eq:expand_Q_Delta}) and (\ref{eq:L(y^2)}).
Our assumption in eq.\ (\ref{eq:assume_alpha&beta}) besides these
factors is that the $\{T,\lambda\}$ dependence is factorized and is
given simply by $\prod_n C_n^{T_n}D_n^{\lambda_n}$.
Surprisingly, for this particular form of
$\alpha_{\Omega,Z,\{T,\lambda\}}^{(\sigma)}$ and
$\beta_{\Omega,Z,\{T,-\lambda\}}^{(\sigma)}$, we can explicitly carry
out the summations in eq.\ (\ref{eq:u_Z-1}).
Namely, we substitute eqs.\ (\ref{eq:expand_Q_Delta}),
(\ref{eq:Phi(oxi_n,oeta_n)}), (\ref{eq:L(y^2)})  and
(\ref{eq:assume_alpha&beta}) into $u_{\Omega,Z,A}^{(\sigma)}$ of eq.\
(\ref{eq:u_Z-1}) and carry out the summations over $\{T,\lambda\}$,
$\ell$ in eq.\ (\ref{eq:expand_Q_Delta}), and
$p_n$ for the Laguerre polynomial in eq.\ (\ref{eq:L(y^2)}).
After a tedious but straightforward calculation which is summarized in
Appendix \ref{app:deriv_u_Z}, we obtain
\begin{eqnarray}
&&u_{\Omega,Z,A}^{(\sigma)}\left(\oxi,\oeta,\oxi_n,\oeta_n\right)
=\exp\left(-i\sigma\Omega\oeta-\sum_n\frac{n\pi}{4}\oxi_n^2\right)\nn\\
&&\quad\times
\Half\int_0^\infty\frac{du}{u}
\exp\left\{\frac{i}{2}\mt\oxi\left(u-\frac{1}{u}\right)\right\}\nn\\
&&\quad\times
\Biggl\{
\alpha_\Omega^{(\sigma)} u^{-i\sigma\Omega}
\prod_n \exp\left[
\left(\frac{u}{D_n e^{\oeta_n-\oeta}}
+\frac{D_n e^{\oeta_n-\oeta}}{u}\right)C_n\sqrt{\frac{n\pi}{2}}\oxi_n
e^{\frac{i\pi n\oxi}{\mt}u}
- C_n^2 e^{\frac{2i\pi n\oxi}{\mt}u}\right]\nn\\
&&\quad + \beta_\Omega^{(\sigma)} u^{i\sigma\Omega}
\prod_n \exp\left[
\left(\frac{u}{D_n e^{\oeta_n-\oeta}}
+\frac{D_n e^{\oeta_n-\oeta}}{u}\right)C_n\sqrt{\frac{n\pi}{2}}\oxi_n
e^{-\frac{i\pi n\oxi}{\mt}\frac{1}{u}}
- C_n^2 e^{-\frac{2i\pi n\oxi}{\mt}\frac{1}{u}}\right]
\Biggr\} ,
\label{eq:u_Z-2}
\end{eqnarray}
where the regularization necessary to make the $u$-integration
well-defined will be taken into account later, and
we have omitted the transverse wave function $\phi_A$ for
simplicity.

We have not yet succeeded in giving the full analysis of the wave
function $u_{\Omega,Z,A}^{(\sigma)}$ of eq.\ (\ref{eq:u_Z-2}) for a
general string configuration specified by
$(\oxi,\oeta,\oxi_n,\oeta_n)$.
Here we shall content ourselves with the analysis for a simple
string configuration satisfying
\begin{equation}
\oeta_n=\oeta\ ,
\end{equation}
for all $n=1,2,\cdots$. This implies that we are considering the
string configuration which does not fluctuate in the $\eta$ direction:
\begin{eqnarray}
&&\eta(\sigma)=\oeta=\sigma\mbox{-independent}\ ,\nn\\
&&\xi(\sigma)=\oxi + \sum_n\oxi_n\cos n\sigma\ .
\label{eq:eta_const_string}
\end{eqnarray}
For such a string configuration, let us adopt the following $C_n$ and
$D_n$:
\begin{equation}
C_n=\frac{\mt}{\sqrt{2\pi n}}\cos n\theta\ ,
\quad D_n=-i\ .
\label{eq:C_nD_n}
\end{equation}
Then, making the change of integration variables $u\to 1/u$ for the
$\beta_\Omega^{(\sigma)}$ term, our Rindler mode (\ref{eq:u_Z-2}),
which we denote as $u_{\Omega,\theta,A}^{(\sigma)}$ with the index
$\theta$, is reduced to
\begin{eqnarray}
&&u_{\Omega,\theta,A}^{(\sigma)}(\oxi,\oeta,\oxi_n,\oeta_n=\oeta)
=\nn\\
&&\qquad\qquad
\biggl\{
\alpha_\Omega^{(\sigma)} R\left(\oxi,\oxi_n,\theta\right)
+\beta_\Omega^{(\sigma)} R\left(-\oxi,-\oxi_n,\theta\right)
\biggr\}e^{-i\sigma\Omega\oeta-\sum_n\left(n\pi/4\right)(\oxi_n)^2}\ ,
\label{eq:u_Z-3}
\end{eqnarray}
using $R\left(\oxi,\oxi_n,\theta\right)$ defined by
\begin{eqnarray}
&&R\left(\oxi,\oxi_n,\theta\right)=
\Half\int_0^\infty\frac{du}{u}u^{-i\sigma\Omega}\nn\\
&&\qquad\qquad\times
\exp\Biggl\{
\frac{i}{2}\mt
\left[\oxi + \sum_n\oxi_n\cos n\theta\,
e^{\frac{i\pi n\oxi}{\mt}u}\right]
\left(u-\frac{1}{u}\right)\Biggr\}
F(\oxi,\theta;u)\ ,
\label{eq:R}
\end{eqnarray}
with
\begin{equation}
F(\oxi,\theta;u)
=\left[\left(1-e^{\frac{2i\pi\oxi}{\mt}u}\right)^2
\left(1-e^{2i\theta+\frac{2i\pi\oxi}{\mt}u}\right)
\left(1-e^{-2i\theta+\frac{2i\pi\oxi}{\mt}u}\right)
\right]^{\mt^2/8\pi}\ .
\label{eq:F}
\end{equation}

To make the $u$-integration of eq.\ (\ref{eq:R}) well-defined for any
$(\oxi,\oxi_n)$, we multiply its integrand by
$\exp\left\{-\epsilon\left(u+(1/u)\right)\right\}$ (which corresponds
to the original regularization (\ref{eq:regfac}) for $Q_\nu(x)$),
and in addition replace $\exp\left(\frac{i\pi n\oxi}{\mt}u\right)$ in
eqs.\ (\ref{eq:R}) and (\ref{eq:F}) by the regularized ones,
$\exp\left\{\left(\frac{i\pi{}n\oxi}{\mt}-\epsilon\right)u\right\}$
($n=2$ for eq.\ (\ref{eq:F})), and take the limit $\epsilon\to +0$
after the integration.

Then, as we did for the function $Q_\nu(x)$ in Sec.\ \ref{sec:BC_u-I},
let us consider the ``Wick rotation'' of the $u$-integration of
$R(\oxi,\oxi_n,\theta)$, namely, the deformation of the
$u$-integration contour from the positive real axis to the imaginary
one; $u\to \pm i u$. In the present case, the direction of the ``Wick
rotation'' depends on the signs of
$\oxi=\int_0^\pi \left(d\sigma/\pi\right)\xi(\sigma)$ and
$\xi(\theta)=\oxi+\sum_n\oxi_n\cos n\theta$:
the behavior of the integrand around $u=\infty$ requests the rotation
$u\to i\sgn(\oxi)u$ ($u>0$), while the behavior around $u=0$ requests
$u\to i\sgn\left(\xi(\theta)\right)u$.
Therefore, defining the condition $(r,s)$ ($r=\pm$, $s=\pm$)
concerning the signature of $\oxi$ and $\xi(\theta)$ by
\begin{equation}
(r,s):\ \sgn(\oxi)=r\quad\mbox{and}\quad
\sgn\left(\xi(\theta)\right)=s\ ,
\label{eq:(r,s)}
\end{equation}
we see that a consistent ``Wick rotation'' is possible only for
$(+,+)$ and $(-,-)$. Corresponding to the cases $(\pm,\pm)$, we have
\begin{equation}
u_{\Omega,\theta,A}^{(\sigma)}
\propto
\left(\alpha_\Omega^{(\sigma)}e^{\pm\pi\sigma\Omega/2}
+\beta_\Omega^{(\sigma)}e^{\mp\pi\sigma\Omega/2}\right)
R_E(\pm\oxi,\pm\oxi_n,\theta)\ ,
\label{eq:u_E}
\end{equation}
where $R_E$ is defined by
\begin{eqnarray}
&&R_E\left(\oxi,\oxi_n,\theta\right)=
\Half\int_0^\infty\frac{du}{u}u^{-i\sigma\Omega}\nn\\
&&\qquad\times
\exp\Biggl\{-\Half\mt\left[
\oxi + \sum_n\oxi_n\cos n\theta\,
 e^{-\frac{\pi n\oxi}{\mt}u}\right]
\left(u+\frac{1}{u}\right)\Biggr\}
F_E(\oxi,\theta;u)\ ,
\label{eq:R_E}
\end{eqnarray}
with
\begin{equation}
F_E(\oxi,\theta;u)
=\left[\left(1-e^{-\frac{2\pi\oxi}{\mt}u}\right)^2
\left(1-e^{2i\theta-\frac{2\pi\oxi}{\mt}u}\right)
\left(1-e^{-2i\theta-\frac{2\pi\oxi}{\mt}u}\right)
\right]^{\mt^2/8\pi} .
\label{eq:F_E}
\end{equation}
Therefore, if $\alpha_\Omega^{(\sigma)}$ and $\beta_\Omega^{(\sigma)}$
are related by
\begin{equation}
\beta_\Omega^{(\sigma)}=-e^{-\pi\Omega}\alpha_\Omega^{(\sigma)}\ ,
\end{equation}
the present $u_{\Omega,\theta,A}^{(+)}$ ($u_{\Omega,\theta,A}^{(-)}$)
satisfies the wedge condition that it vanishes when both
the CM coordinate $\oxi$ and the string point $\xi(\theta)$ at a
particular parameter value are negative (positive).
Our $u_{\Omega,\theta,A}^{(\pm)}$ does not vanish
when $\oxi$ and $\xi(\theta)$ are in different wedges,

The analysis in this section shows that, by summing over
$\{T,\lambda\}$, we can construct string Rindler modes with
wedge conditions different from the one for
$u_{\Omega\{T,\lambda\}A}^{(\sigma)}$ of Sec.\ 3.
Although it is not clear whether the present wedge condition
itself has any physical significance, our result is expected to give
a clue to the construction of the string Rindler modes with more
natural wedge conditions.
As regards the analysis in this section, there are a number of points
to be clarified.
First, we have restricted our consideration only to the
string configurations with $\eta(\sigma)=\sigma\mbox{-independent}$.
For a general string configurations with a non-trivial $\eta(\sigma)$,
our Rindler mode $u_{\Omega,\theta,A}^{(\sigma)}$ does not seem to
obey a simple wedge condition. It is also an open question
whether we can construct a complete basis of Rindler modes containing
the present $u_{\Omega,\theta,A}^{(\sigma)}$.\footnote{
The inner product of our
$u_{\Omega,\theta,A}^{(\sigma)}$
is proportional to
$$
\left(u_{\Omega,\theta,A}^{(\sigma)},
u_{\Omega',\theta',A'}^{(\sigma')}\right)
\propto
\delta_{\sigma,\sigma'}\delta\left(\Omega-\Omega'\right)\eta_{AA'}
\abs{\cos\theta-\cos\theta'}^{\mt^2/2\pi}\ .
$$
}

\section{Remaining problems}

In this paper we have learned many things about the quantization of SFT
in the Rindler space-time.
However, there remain a number of problems to be clarified before
reaching the complete understanding.
In the following we shall summarize these problems, although some of
them were already mentioned in the text.

The first problem is concerned with the wedge condition of the string
Rindler mode, or the division of the Hilbert space of states.
In Secs.\ 3 and 4, we adopted the wedge condition
(\ref{eq:BC_u}) specified by the Minkowski CM coordinate $x^\mu$.
This choice, however, is mainly due to the technical reason that our
string Rindler mode is constructed using the components of the
Minkowski modes.
Another wedge condition specified by the Rindler CM coordinate
$\xi_0$ should also be considered seriously.
Furthermore, as stated in Sec.\ 1, this problem of
the wedge condition of the Rindler mode is related to
the causality problem in SFT, which has been recently investigated
in the Minkowski space-time using the light-cone gauge SFT
\cite{Martinec,Love,LPSTU}.
The causality relation in SFT on the
Rindler space-time needs to be studied in detail.
In particular, it is crucial to understand the causality structure
for the string which crosses the origin (i.e., the string (B)
depicted in Fig.\ \ref{fig:wedge}) for solving this problem.
For this purpose, the techniques used in constructing the Rindler
modes with an interesting wedge condition in Sec.\ 5 may be
helpful.
Anyway, it is an open question to determine the causality structure
of SFT in the Rindler space-time and at the same time
to find the most natural wedge condition for the Rindler modes.

The second problem we have to clarify is to present the physical
observable quantities in the Rindler quantization of SFT.
In Sec.\ 4.2, we calculated the Minkowski vacuum expectation value of
the Rindler Hamiltonian, eq.\ (\ref{eq:vac_MH_Rvac_M}).
As stated there, we need a more detailed analysis by introducing the
regularization for $\delta(\Omega=0)$.
In Ref.\ \cite{SusskindUglum}, it was argued that the black hole
entropy per unit area is finite to all orders in superstring
perturbation theory, though it is divergent in the case of a free
scalar field.
It is interesting to examine this observation
from the viewpoint of SFT which we have presented here.
 In addition, it would be necessary to study whether we could devise
the ``detector'' in string theory as in particle field theory
\cite{BD,bible}.

As our third problem, we should examine the inner product for the
string modes more precisely. In this paper, we have assumed that the
Rindler inner product (\ref{eq:innprod_R}) equals the Minkowski inner
product (\ref{eq:innprod_M}), and this was crucial for our analysis in
this paper. Although the equality of the two inner products holds
if certain surface terms vanish as is explained in Appendix A,
mathematically detailed examination is necessary.

\vspace{1.5cm}
\centerline{\Large\bf Appendix}
\appendix

\section{Inner product for string wave functions}
\label{app:innprod_R}

In this Appendix we shall explain the conserved inner product for the
string wave functions.
Let us consider SFT (\ref{eq:S_M}) described by a string coordinate
$Y^\mu(\sigma)$
related to the Minkowski one $X^\mu(\sigma)$ by a coordinate
transformation:
\begin{equation}
X^\mu(\sigma)=f^\mu\left(Y(\sigma)\right)\ .
\label{eq:X=f(Y)}
\end{equation}
The SFT action rewritten in terms of the new string coordinate
$Y^\mu(\sigma)$ reads
\begin{eqnarray}
S=-\Half\int\calD Y^\mu(\sigma)\,\calD(\mbox{ghosts})
\prod_{\sigma'}\sqrt{-G\left(Y(\sigma')\right)}
\ \Phi L\Phi\ ,
\label{eq:S_Y}
\end{eqnarray}
with $L$ (\ref{eq:L_M}) also reexpressed using $Y^\mu$ as
\begin{eqnarray}
&&L=\pi\int_0^{\pi} d\sigma
\left\{
-\frac{1}{\sqrt{-G}}
\hanbi{}{Y^{\mu}}\left(\sqrt{-G}G^{\mu\nu}
\hanbi{}{Y^{\nu}}\right)
+G_{\mu\nu}Y^{\prime\mu}Y^{\prime\nu}
+(\mbox{ghost part})
\right\}\ .
\label{eq:L_R}
\end{eqnarray}
In eqs.\ (\ref{eq:S_Y}) and (\ref{eq:L_R}), the metric
$G_{\mu\nu}\left(Y(\sigma)\right)$ is given by
\begin{equation}
G_{\mu\nu}\left(Y(\sigma)\right)
=\left.\eta_{\alpha\beta}\henbi{f^\alpha(y)}{y^\mu}
\henbi{f^\beta(y)}{y^\nu}\right\vert_{y=Y(\sigma)}\ ,
\label{eq:G_munu}
\end{equation}
and $G\equiv \det G_{\mu\nu}$.
Functional integration by parts gives another expression of $S$,
\begin{eqnarray}
&&S=-\frac{\pi}{2}\int\calD Y^\mu(\sigma)\,\calD(\mbox{ghosts})
\prod_{\sigma'}\sqrt{-G\Bigl(Y(\sigma')\Bigr)}\nn\\
&&\qquad\times
\int_0^\pi d\sigma\Biggl\{
G^{\mu\nu}\Bigl(Y(\sigma)\Bigr)\hanbi{\Phi}{Y^\mu(\sigma)}
\hanbi{\Phi}{Y^\nu(\sigma)}
+G_{\mu\nu}\Bigl(Y(\sigma)\Bigr)Y'^\mu(\sigma)Y'^\nu(\sigma)
\Phi^2\nn\\
&&\hspace*{7cm}
+(\mbox{ghost part})\Biggr\}\ .
\label{eq:S_by_Y}
\end{eqnarray}
We assume that a mathematically awkward expression,
\begin{equation}
\prod_\sigma \sqrt{-G(\sigma)}=\exp\left\{
\delta(\sigma=0)\int_0^\pi d\sigma \ln \sqrt{-G(\sigma)}\right\}\ ,
\label{eq:prod_sigma}
\end{equation}
is defined by means of a suitable regularization, which allows
formal and naive functional manipulations.

Let $\Phi_1$ and $\Phi_2$ be string wave functions satisfying
the equation of motion, $L\Phi_{1,2}=0$, the current $J_{12}^\mu$
defined by
\begin{equation}
J_{12}^\mu\left(Y;\sigma\right)
=G^{\mu\nu}\left(Y(\sigma)\right)\,\Phi_1^*
\sayuu{\hanbi{}{Y^\nu(\sigma)}}\Phi_2\ ,
\label{eq:J12}
\end{equation}
is conserved in the following sense:
\begin{equation}
\int_0^{\pi}d\sigma \frac{1}{\sqrt{-G(\sigma)}}
\hanbi{}{Y^\mu(\sigma)}\left(
\sqrt{-G(\sigma)}\,J_{12}^\mu(Y;\sigma)\right)=0\ ,
\label{eq:DJ=0}
\end{equation}
with $G(\sigma)$ short for $G\left(Y(\sigma)\right)$.
Integrating eq.\ (\ref{eq:DJ=0}) over some (functional) region $\calV$,
we get
\begin{eqnarray}
&&0=
\int_{\calV}\calD Y^\mu\calD(\mbox{ghost})
\prod_{\sigma'}\sqrt{-G(\sigma')}
\times\mbox{eq.\ (\ref{eq:DJ=0})}\nn\\
&&\phantom{0}=
\int_{\calV}\calD Y^\mu\calD(\mbox{ghost})
\int_0^{\pi}d\sigma\hanbi{}{Y^\nu(\sigma)}\left(
\prod_{\sigma'}\sqrt{-G(\sigma')}\cdot
J_{12}^\nu(Y;\sigma)\right)\ ,
\label{eq:surface}
\end{eqnarray}
where the easiest way to understand the last equality is to regard
$\prod_{\sigma'}$ as a discrete product.

In the case of the Rindler coordinate,
\begin{equation}
Y^\mu(\sigma)=\left(\eta(\sigma), \xi(\sigma), X^\bot(\sigma)
\right)\ ,
\label{eq:Y_R}
\end{equation}
we have
\begin{equation}G_{\eta\eta}=-\left(\xi(\sigma)\right)^2,
\quad G_{\xi\xi}=1,
\quad G_{\bot\bot}={\rm diag}(1,\cdots,1),
\quad \mbox{others}=0 ,
\label{eq:G_munu_R}
\end{equation}
and
\begin{equation}
\sqrt{-G}\equiv \sqrt{-\det G_{\mu\nu}}=\abs{\xi(\sigma)}\ .
\label{eq:G_R}
\end{equation}
Taking as $\calV$ in eq.\ (\ref{eq:surface}) the region bounded by two
different values of $\eta_0$, and
assuming that the surface terms vanish except for the one for the
zero-mode $\eta_0$, we see that the Rindler inner product defined by
\begin{eqnarray}
(\Phi_1,\Phi_2)_{\rm R}=
i\int\frac{\calD\eta\calD\xi\calD X_{\bot}
\calD(\mbox{ghost})}{d\eta_0}
\prod_{\sigma'}\abs{\xi(\sigma')}\int_0^\pi\! d\sigma
\frac{1}{\xi(\sigma)^2}\,\left.
\Phi_1^*\sayuu{\hanbi{}{\eta(\sigma)}}\Phi_2
\right|_{\eta_0={\rm const.}} ,
\label{eq:innprod_R}
\end{eqnarray}
is independent of the choice of $\eta_0$.
On the other hand, taking as $Y^\mu(\sigma)$ the original Minkowski
string coordinate $X^\mu(\sigma)$, eq.\ (\ref{eq:surface}) leads to the
conserved Minkowski inner product (\ref{eq:innprod_M}).
Finally, by considering in eq.\ (\ref{eq:surface}) the region $\calV$
bounded on the one hand by $\eta_0=\mbox{const}$ and on the other by
$x^0=\mbox{const}$ and assuming again that the other surface terms
vanish, it follows that the Minkowski inner product
(\ref{eq:innprod_M}) and the Rindler one (\ref{eq:innprod_R}) are
equal.

\section{Two other expressions of $\Phi_{T_n,\lambda_n}$}
\label{app:Two_other}

In Sec.\ 3, the oscillator part wave function $\Phi_{T_n,\lambda_n}$
(\ref{eq:Phi_T_nlambda_n}) of the Rindler mode
$u_{\Omega\{T,\lambda\}A}^{(\sigma)}$ was given as a linear
combination of the Minkowski oscillator mode $\Psi_{M_n,N_n}$
(\ref{eq:Psi^n_MN}).
In this Appendix, we present two other expressions of
$\Phi_{T_n,\lambda_n}$ which are equivalent to eq.\
(\ref{eq:Phi_T_nlambda_n}).

In the following we use the coordinates $(\oxi_n,\oeta_n)$ ($n\ge 1$)
of eq.\ (\ref{eq:oxi_n-oeta_n}), or equivalently,
\begin{equation}
x^{\pm}_n\equiv x^1_n \pm x^0_n=\oxi_n \exp\left(\pm\oeta_n\right)\ .
\label{eq:x_n^pm}
\end{equation}
Note that
\begin{equation}
\henbi{}{\oeta_n}= x^+_n\henbi{}{x^+_n} - x^-_n\henbi{}{x^-_n}
=x^0_n\henbi{}{x^1_n} + x^1_n\henbi{}{x^0_n}\ .
\label{eq:p/ph_n}
\end{equation}

\noindent
\underline{Fock space expression}

The first expression is based on the Fock space representation of
the oscillator modes.
Let us define the creation/annihilation operators $\alpha^\mu_n$
($n=\pm 1, \pm 2,\cdots$) in terms of $x^\mu_n$ and $\p/\p x^\mu_n$ of
(\ref{eq:X_mode}) and (\ref{eq:d/dX_mode}) as
\begin{equation}
\alpha^\mu_n=-\frac{i}{\sqrt{\pi}}\left(
\frac{n\pi}{2} x^\mu_{|n|} +
\eta^{\mu\nu}\henbi{}{x^\nu_{|n|}}\right)\ ,
\label{eq:a_n}
\end{equation}
which satisfies the commutation relation,
\begin{equation}
\left[\alpha^\mu_n,\, \alpha^\nu_m\right]
=n\delta_{n+m,0}\eta^{\mu\nu}\ .
\label{eq:CCR_a_n}
\end{equation}
The $x_n^\mu$-oscillator part of the $(\mbox{mass})^2$ operator in
eq.\ (\ref{eq:L_M}) is given in terms of $\alpha_n^\mu$ as
\begin{equation}
(\mbox{mass})^2_{n,\mu}\equiv2\eta^{\mu\mu}\left\{
-\henbi{}{x_n^\mu}\henbi{}{x_n^\mu}
+\left(\frac{n\pi}{2}\right)^2 x_n^\mu x_n^\mu\right\}
=2\pi \left(\eta^{\mu\mu}\alpha_{-n}^\mu\alpha_n^\mu
+ \frac{n}{2}\right)\ .
\end{equation}
Then the light-cone component $\alpha^\pm_n$,
\begin{equation}
\alpha^{\pm}_n\equiv\frac{1}{\sqrt{2}}\left(
\alpha^1_n\pm \alpha^0_n\right)
=-i\sqrt{\frac{2}{\pi}}\left(
\frac{n\pi}{4} x^{\pm}_{|n|} + \henbi{}{x^{\mp}_{|n|}}\right) ,
\end{equation}
satisfies the commutation relations,
\begin{equation}
\left[\alpha^\pm_n,\,\alpha^\mp_m\right]
=n\delta_{n+m,0}\ ,
\quad
\left[\alpha^\pm_n,\,\alpha^\pm_m\right]=0\ .
\end{equation}
We also need the commutation relation between $\p/\p\oeta_n$ and
$\alpha^\pm_n$:
\begin{equation}
\left[\henbi{}{\oeta_n},\, \alpha^\pm_m \right]
=\pm\, \delta_{n,|m|}\,\alpha^\pm_m\ .
\label{eq:[d/dh,alpha]}
\end{equation}

Now recall that the wave function $\Phi_{T_n,\lambda_n}$ is
characterized by the two conditions:
\begin{eqnarray}
\henbi{}{\oeta_n}\Phi_{T_n,\lambda_n}=\lambda_n\Phi_{T_n,\lambda_n}
\label{eq:d/dhPhi=lambdaPhi}\ ,\\
\widehat T_n\Phi_{T_n,\lambda_n}
=T_n\Phi_{T_n,\lambda_n}\ ,
\label{eq:hatTPhi=TPhi}
\end{eqnarray}
where the first equation (\ref{eq:d/dhPhi=lambdaPhi}) is nothing but
eq.\ (\ref{eq:alpha_nPsi_MnNn}),
and $\widehat T_n$ is the level number operator of the $n$-th mode
of the $\mu=0$ and $1$ oscillators:
\begin{equation}
n\widehat T_n=\alpha_{-n}^1\alpha_n^1 - \alpha_{-n}^0\alpha_n^0
=\alpha_{-n}^+\alpha_n^- + \alpha_{-n}^-\alpha_n^+ \ .
\label{eq:hatT_n}
\end{equation}
The creation operators $\alpha_{-n}^\pm$ ($n\ge 1$) shift $\lambda_n$ by
$\pm 1$ (see eq.\ (\ref{eq:[d/dh,alpha]})) and they both raise $T_n$ by
one; namely, we have
\begin{equation}
\alpha^\pm_{-n}\Phi_{T_n,\lambda_n}\propto
\Phi_{T_n+1,\lambda_n \pm 1}\quad (n\ge 1)\ .
\label{eq:a^pmPhi}
\end{equation}
The wave function $\Phi_{T_n,\lambda_n}$
with the normalization condition
\begin{equation}
\left(\Phi_{T_n,\lambda_n},\Phi_{T'_n,\lambda'_n}\right)
=\delta_{T_n,T'_n}\delta_{\lambda_n+\lambda'_n,0}\ ,
\end{equation}
is therefore given as
\begin{equation}
\Phi_{T_n,\lambda_n}=\frac{1}{\sqrt{N_+! N_-!\, n^{N_+ + N_-}}}
\left(\alpha^+_{-n}\right)^{N_+}
\left(\alpha^-_{-n}\right)^{N_-}
\Phi_{0,0}\ ,
\label{eq:aa0}
\end{equation}
where $\Phi_{0,0}$ is the Fock space vacuum satisfying
$\alpha_n^\pm\Phi_{0,0}=0$ ($n\ge 1$), and $N_\pm$ is given by
\begin{equation}
N_{\pm}=\frac{1}{2}\left(T_n \pm \lambda_n\right)\ .
\end{equation}
In this construction, it is evident that $\lambda_n$ should lie in the
range of (\ref{eq:allowedlambda_n}) since $N_\pm$ are non-negative
integers.

\noindent
\underline{Expression by the coordinate $(\oxi_n, \oeta_n)$}

Next we shall express $\Phi_{T_n,\lambda_n}$ as a function of the
variables $(\oxi_n,\oeta_n)$ of eq.\ (\ref{eq:oxi_n-oeta_n}).
The operator $\widehat T_n$ in (\ref{eq:hatTPhi=TPhi}) is expressed in
terms of $(\oxi_n,\oeta_n)$ as follows:
\begin{eqnarray}
n\pi\left(\widehat T_n+1\right)=
-\left(\henbi{}{\oxi_n}\right)^2 -\frac{1}{\oxi_n}\henbi{}{\oxi_n}
+\frac{1}{\oxi_n^2}\left(\henbi{}{\oeta_n}\right)^2
+\left(\frac{n\pi}{2}\right)^2\oxi_n^2\ ,
\end{eqnarray}
where $1=(1/2)\times 2$ on the LHS is the zero-point energy of
the two oscillators.
The condition (\ref{eq:hatTPhi=TPhi}) is now reduced to the
differential equation for the variable $\oxi_n$ (the $\oeta_n$
dependence is fixed by another condition (\ref{eq:d/dhPhi=lambdaPhi})
to be $\Phi_{T_n,\lambda_n}\propto \exp\left(\lambda_n\oeta_n\right)$).
This differential equation has a solution regular at $\oxi_n=0$
only when $T_n$ is a non-negative integer and $\lambda_n$ is in the
range of (\ref{eq:allowedlambda_n}).
For such $(T_n,\lambda_n)$, the normalized solution
is given by
\begin{equation}
\Phi_{T_n,\lambda_n}\left(\oxi_n,\oeta_n\right)=
\sqrt{\frac{
\frac{\ds n}{\ds 2}
\left(\frac{\ds T_n-\abs{\lambda_n}}{\ds 2}\right)!}
{\left(\frac{\ds T_n+\abs{\lambda_n}}{\ds 2}\right)!}}
\left(\frac{n\pi}{2}\oxi_n^2\right)^{\abs{\lambda_n}/2}
L_{\frac{1}{2}\left(T_n-\abs{\lambda_n}\right)}^{(\abs{\lambda_n})}
\left(\frac{n\pi}{2}\oxi_n^2\right)\exp\left(
-\frac{n\pi}{4}\oxi_n^2 + \lambda_n \oeta_n\right)\ ,
\label{eq:Phi(oxi_n,oeta_n)}
\end{equation}
where $L_m^{(\alpha)}(x)$ is the Laguerre polynomial:
\begin{equation}
L_m^{(\alpha)}(x)=\sum_{r=0}^m(-)^r\pmatrix{m+\alpha\cr m-r}
\frac{x^r}{r!}\ .
\label{eq:Laguerre}
\end{equation}
Namely,
$\Phi_{T_n,\lambda_n}\exp\left(\frac{n\pi}{4}\oxi_n^2
- \lambda_n \oeta_n\right)$
is a polynomial of $\oxi_n$ of the form,
\begin{equation}
\oxi_n^{T_n} + \oxi_n^{T_n-2} + \cdots
+ \oxi_n^{\abs{\lambda_n}+2} + \oxi_n^{\abs{\lambda_n}}\ ,
\end{equation}
up to the coefficient of each term.
Corresponding to the fact that the inner product integration over
$x_n^0$ should be carried out in the pure-imaginary direction (see
eq.\ (\ref{eq:innprod_M})), the inner product integration over the
present variables $(\oxi_n,\oeta_n)$ should be defined by regarding
$\oxi_n$ as the radius variable and $i\oeta_n$ as the (real) angle
variable.
Therefore, $\Phi_{T_n,\lambda_n}$ is normalized in the following
sense:
\begin{equation}
\int_0^\infty\! \oxi_n d\oxi_n\int_0^{2\pi}\!d\left(i\oeta_n\right)
\Phi^*_{T'_n,\lambda'_n}(\oxi_n,\oeta_n)
\Phi_{T_n,\lambda_n}(\oxi_n,\oeta_n)
=\delta_{T'_n,T_n}\delta_{\lambda_n+\lambda'_n,0}\ ,
\label{eq:innprod_oxioeta}
\end{equation}
where the complex conjugation of
$\Phi_{T'_n,\lambda'_n}(\oxi_n,\oeta_n)$
should be done by regarding $\oeta_n$ as a real variable.
The integration of (\ref{eq:innprod_oxioeta}) can be carried out
by using the formula for the Laguerre polynomial:
\begin{equation}
\int_0^\infty\! dx\,e^{-x}x^\alpha L_m^{(\alpha)}(x)L_n^{(\alpha)}(x)
=\frac{\Gamma(\alpha+m+1)}{m!}\delta_{m,n}\ .
\end{equation}

\section{BRST invariance of the Rindler vacuum}
\label{app:QBket0_R=0}

In this appendix we show that the Rindler vacuum state $\rket{0}$
defined by eq.\ (\ref{eq:vac_R}) is BRST invariant and hence is a
physical state. Namely, we show that \cite{KugoOjima}
\begin{equation}
\bQB\rket{0}=0\ ,
\label{eq:bQBket0_R=0}
\end{equation}
where $\bQB$ is the BRST operator of the free SFT generating the BRST
transformation $\dB$:
\begin{equation}
\left[i\bQB,\Phi\right]=\dB\Phi\equiv \tQB\Phi\ .
\label{eq:def_bQB}
\end{equation}
In eq.\ (\ref{eq:def_bQB}), $\tQB$ is the BRST charge in the first
quantized string theory with terms containing the ghost zero-modes
$\ac_0$ and $c_0=\p/\p\ac_0$ omitted \cite{KatoOgawa,HIKKOopen}:
\begin{equation}
\tQB=-\Half\sum_{n,m}c_{-n}\Bigl(
\eta_{\mu\nu}\alpha_{n-m}^\mu\alpha_{m}^\nu
+(n+m)\ac_{n-m}c_m\Bigr)\Big|_{c_0=\ac_0=0}\ ,
\label{eq:tQB}
\end{equation}
where $\alpha_0^\mu=-i\p/\p x_\mu(=k^\mu$ when $\alpha_0^\mu$ acts on
$U_k$ (\ref{eq:U_k})).
$\bQB$ and $\tQB$ should not be confused: the former is the operator
in the second quantized string theory (SFT) while the
latter is the one in the first quantized string theory.

To show the BRST invariance of $\rket{0}$,
it is sufficient to show that
\begin{equation}
\left(u_{\Omega\{T,\lambda\}B}^{(\sigma)},
\tQB u_{\Omega'\{T',\lambda'\}A'}^{(\sigma')*}\right)
=
\left(u_{\Omega\{T,\lambda\}B}^{(\sigma)*},
\tQB u_{\Omega'\{T',\lambda'\}A'}^{(\sigma')}\right)
=0\ .
\label{eq:cond_QBvac_R=0}
\end{equation}
To see why eq.\ (\ref{eq:cond_QBvac_R=0}) is sufficient, note first that
\begin{eqnarray}
&&\dB b_{\Omega\{T,\lambda\}}^{(\sigma)A}=
(-)^{|A|}\sum_B\eta^{AB}\left(
u_{\Omega\{T,\lambda\}B}^{(\sigma)},\tQB\Phi\right)\nn\\
&&\phantom{\dB b_{\Omega\{T,\lambda\}}^{(\sigma)A}}
=(-)^{|A|}\sum_B\eta^{AB}\sum_{\sigma'}\int d\Omega'
\sum_{{\scriptstyle\{T',\lambda'\}\atop \scriptstyle A'}}
\biggl\{
\left(u_{\Omega\{T,\lambda\}B}^{(\sigma)},
\tQB u_{\Omega'\{T',\lambda'\}A'}^{(\sigma')}\right)
b_{\Omega'\{T',\lambda'\}}^{(\sigma')A'}\nn\\
&&\qquad\qquad\qquad +(-)^{|A'|}
\left(u_{\Omega\{T,\lambda\}B}^{(\sigma)},
\tQB u_{\Omega'\{T',\lambda'\}A'}^{(\sigma')*}\right)
b_{\Omega'\{T',\lambda'\}}^{(\sigma')A'\dagger}
\biggr\}\ ,
\end{eqnarray}
and a similar equation for
$\dB b_{\Omega\{T,\lambda\}}^{(\sigma)A\dagger}$.
Therefore, if eq.\ (\ref{eq:cond_QBvac_R=0}) is satisfied,
$\bQB$ is expressed in terms of the Rindler creation/annihilation
operators as
\begin{equation}
\bQB=i\sum_{\sigma,\sigma'}\int\! d\Omega\int\! d\Omega'
\sum_{{\scriptstyle \{T,\lambda\}\atop \scriptstyle A}}
\sum_{{\scriptstyle\{T',\lambda'\}\atop\scriptstyle A'}}
b_{\Omega\{T,\lambda\}}^{(\sigma)A\dagger}
\left(u_{\Omega\{T,\lambda\}A}^{(\sigma)},
\tQB u_{\Omega'\{T',\lambda'\}A'}^{(\sigma')}\right)
b_{\Omega'\{T',\lambda'\}}^{(\sigma')A'}\ ,
\label{eq:bQB_R}
\end{equation}
which implies eq.\ (\ref{eq:bQBket0_R=0}).
There is no ordering ambiguity in (\ref{eq:bQB_R}) since
$b^\dagger$ and $b$ there have opposite statistics.

In the following we show the vanishing of the second term of eq.\
(\ref{eq:cond_QBvac_R=0}) (the first term is in fact equal to the
second term due to the hermiticity of $\tQB$).
For this purpose we separate $\tQB$ (\ref{eq:tQB}) into three parts:
\begin{equation}
\tQB=\til{Q}_\bot +\til{Q}_{ck\alpha}+ \til{Q}_{c\alpha\alpha}\ ,
\end{equation}
where $\til{Q}_\bot$ is the part of $\tQB$ which contains no
$\alpha^\mu_n$ with $\mu=0, 1$,
and the other two terms are defined by
\begin{eqnarray}
&&\til{Q}_{ck\alpha}=-\Half
\sum_{n\ne 0} c_{-n}\left(\alpha^+_{n}k^- +\alpha^-_{n}k^+\right)\ ,
\label{eq:tQB_cka}\\
&&\til{Q}_{c\alpha\alpha}
=-\sum_{n,m}{}' c_{-n}\alpha^+_{n-m}\alpha^-_{m}\ ,
\label{eq:tQB_caa}
\end{eqnarray}
using the light-cone components of Appendix \ref{app:Two_other}.
Note that the summation $\sum'_{n,m}$ in eq.\ (\ref{eq:tQB_caa}) is
subject to the constraint $nm(n-m)\ne 0$. In eq.\ (\ref{eq:tQB_cka})
we have replaced $\alpha^\pm_0$ by $k^\pm$. This is valid when
$\til{Q}_{ck\alpha}$ acts on $U_k$ (\ref{eq:U_k}).

Recalling the derivation of (\ref{eq:(u*,u)}), it is evident that
the condition (\ref{eq:cond_QBvac_R=0}) is satisfied for the
$\til{Q}_\bot$ part since the effect of the insertion of
$\til{Q}_\bot$ is only to change $\left(\phi_A^*,\phi_{A'}\right)$ in
(\ref{eq:(u*,u)}) into $\left(\phi_A^*,\til{Q}_\bot\phi_{A'}\right)$.

Next, let us consider the contribution of the $\til{Q}_{ck\alpha}$
term (\ref{eq:tQB_cka}).
Owing to eq.\ (\ref{eq:a^pmPhi}), the inner product of
(\ref{eq:cond_QBvac_R=0}) vanishes unless
\begin{equation}
\sum_n\left(\lambda_n+\lambda'_n\right)=\mp 1\ ,
\label{eq:lambda+lambda'}
\end{equation}
with $\mp$ corresponding to the $\alpha_n^\pm k^\mp$ term of
$\til{Q}_{ck\alpha}$.
In addition, note that under the change of variables $k=\mu\sinh y$ we
have
\begin{equation}
k^\pm=\frac{1}{\sqrt{2}}\left(k\pm\omega_k\right)
=\pm\frac{\mu}{\sqrt{2}}e^{\pm y}\ .
\label{eq:k^pm}
\end{equation}
Therefore, the $k$-integration relevant to the inner product of
(\ref{eq:cond_QBvac_R=0}) with $\til{Q}_{ck\alpha}$ is proportional to
(c.f., eq.\ (\ref{eq:k-int}))
\begin{eqnarray}
&&\int_{-\infty}^\infty\frac{dk}{2\pi\omega_k}k^\mp
\left(\frac{\omega_k+k}{\omega_k-k}\right)^{\!
-i\left(\sigma\Omega+\sigma'\Omega'\right)/2
\pm 1/2}\nn\\
&&\propto\int_{-\infty}^\infty\! \frac{dy}{2\pi}\,
e^{i\left(\sigma\Omega+\sigma'\Omega'\right)y}
=\delta\left(\sigma\Omega+\sigma'\Omega'\right)
=\delta_{\sigma,-\sigma'}\delta(\Omega-\Omega')\ .
\end{eqnarray}
Namely, the contributions of $k^\pm$ and the extra exponent
$-\sum_n\left(\lambda_n+\lambda'_n\right)/2=\pm 1/2$
of $(\omega_k +k)/(\omega_k -k)$ just cancel to give the same
integration as without $\til{Q}_{ck\alpha}$.
Since the $k$-integration implies $\Omega=\Omega'$, the present matrix
element of $\til{Q}_{ck\alpha}$ is proportional to
\begin{equation}
\beta_{\Omega\{T,\lambda\}}^{(\sigma)}
\alpha_{\Omega\{T',\lambda'\}}^{(-\sigma)}
+\alpha_{\Omega\{T,\lambda\}}^{(\sigma)}
\beta_{\Omega\{T',\lambda'\}}^{(-\sigma)}\ .
\label{eq:ba+ab}
\end{equation}
Note that the two terms of eq. (\ref{eq:ba+ab}) have relatively the
same sign contrary to the case of eq.\ (\ref{eq:(u*,u)}):
this is due to the fact that $\alpha_0^\pm U_k^*=-k^\pm U_k^*$
while $\alpha_0^\pm U_k =k^\mu U_k$.
Using the form of $\alpha_{\Omega\{T,\lambda\}}^{(\sigma)}$
and $\beta_{\Omega\{T,\lambda\}}^{(\sigma)}$ given by eq.\
(\ref{eq:alphabeta_final}) and the constraint
(\ref{eq:lambda+lambda'}), we see that the contribution of the
$\til{Q}_{ck\alpha}$ term to (\ref{eq:cond_QBvac_R=0}) vanishes.

Finally, let us consider the contribution from
$\til{Q}_{c\alpha\alpha}$ (\ref{eq:tQB_caa}).
Since we have $\alpha^+_{n-m}\alpha^-_n\Phi_{\{T,\lambda\}}\propto
\Phi_{\{\til{T},\til{\lambda}\}}$ with
$\sum_n\til{\lambda}_n=\sum_n\lambda_n$,
the $k$-integration is again reduced to
$\delta_{\sigma,-\sigma'}\delta(\Omega-\Omega')$.
In this case the matrix element is proportional to
$\beta_{\Omega\{T,\lambda\}}^{(\sigma)}
\alpha_{\Omega\{T',\lambda'\}}^{(-\sigma)}
-\alpha_{\Omega\{T,\lambda\}}^{(\sigma)}
\beta_{\Omega\{T',\lambda'\}}^{(-\sigma)}$, which
is also seen to vanish.
This completes the proof of the BRST invariance of the Rindler vacuum.

Quite similarly but much more easily, we see that
the BRST operator $\bQB$ is expressed in terms of the Minkowski
creation/annihilation operator as
\begin{equation}
\bQB=i\int\! dk\int\! dk'\sum_{\{M,N\},A}\sum_{\{M',N'\},A'}
a_{k\{M,N\}}^{A\dagger}\left(U_{k\{M,N\}A},
\tQB U_{k'\{M',N'\}A'}\right)a_{k'\{M',N'\}}^{A}\ ,
\label{eq:bQB_M}
\end{equation}
which implies the BRST invariance of the Minkowski vacuum,
\begin{equation}
\bQB\mket{0}=0\ .
\end{equation}

\section{Derivation of eq.\ (\protect\ref{eq:u_Z-2})}
\label{app:deriv_u_Z}

In this Appendix, we summarize the derivation
of eq.\ (\ref{eq:u_Z-2}).
What we have to do is to substitute eqs.\ (\ref{eq:expand_Q_Delta}),
(\ref{eq:Phi(oxi_n,oeta_n)}), (\ref{eq:L(y^2)})  and
(\ref{eq:assume_alpha&beta}) into $u_{\Omega,Z,A}^{(\sigma)}$ of eq.\
(\ref{eq:u_Z-1}) and carry out the summations over $(T_n,\lambda_n)$,
$\ell$ in eq.\ (\ref{eq:expand_Q_Delta}), and
$p_n$ for the Laguerre polynomial in eq.\ (\ref{eq:L(y^2)}).
For this purpose it is convenient to make a change of summation
variables from $(T_n,\lambda_n, p_n)$ to $(L_n,m_n,p_n)$ defined by
\begin{eqnarray}
&&L_n=T_n-2p_n\ ,\nn\\
&&m_n=\Half\left(T_n -\lambda_n\right)-p_n
=\Half\left(L_n - \lambda_n\right)\ .
\label{eq:L&m}
\end{eqnarray}
The range of the new summation variables $(T_n,\lambda_n, p_n)$ is
\begin{equation}
0\le m_n\le L_n,\quad p_n\ge 0 \ .
\label{eq:range_Lmp}
\end{equation}
Then using $y_n \equiv \sqrt{n\pi/2}\oxi_n$ instead of $\oxi_n$, we
have
\begin{eqnarray}
&&
\exp\left(i\sigma\Omega\oeta+\sum_n\frac{n\pi}{4}\oxi_n^2\right)\cdot
u_{\Omega,Z,A}^{(\sigma)}\left(\oxi,\oeta,\oxi_n,\oeta_n\right)\nn\\
&&=\prod_n
\sum_{T_n=0}^\infty\sum_{\lambda_n}\sum_{p_n=0}^\infty
\frac{(-)^{p_n}}{p_n!\left(T_n-2p_n\right)!}
\pmatrix{T_n - 2p_n\cr (T_n-\lambda_n)/2-p_n}y_n^{T_n-2p_n}
\left(D_n e^{\oeta_n-\oeta}\right)^{\lambda_n}C_n^{T_n}\nn\\
&&\quad\times\sum_{\ell=0}^\infty\frac{1}{\ell!}
\left(\frac{i\pi}{\mt}\sum_n nT_n\oxi\right)^\ell
\left(
\alpha^{(\sigma)}_\Omega
Q_{i\sigma\Omega+\sum_n\lambda_n-\ell}\left(\mt\oxi\right)
+ (-)^\ell
\beta^{(\sigma)}_\Omega
Q_{-i\sigma\Omega+\sum_n\lambda_n+\ell}\left(\mt\oxi\right)
\right)\nn\\
&&=\Half\int_0^\infty\frac{du}{u}
\exp\left\{\frac{i}{2}\mt\oxi\left(u-\frac{1}{u}\right)
\right\}\nn\\
&&\quad\times
\prod_n\sum_{L_n=0}^\infty\sum_{m_n=0}^{L_n}\sum_{p_n=0}^\infty
\frac{(-)^{p_n}}{p_n!L_n!}
\pmatrix{L_n\cr m_n}y_n^{L_n}
\left(\frac{D_n e^{\oeta_n-\oeta}}{u}\right)^{L_n-2m_n}C_n^{L_n+2p_n}
\nn\\
&&\quad\times\sum_{\ell=0}^\infty \frac{1}{\ell!}
\left(\frac{i\pi}{\mt}\sum_n n\left(L_n+2p_n\right)\oxi\right)^\ell
\left(
\alpha^{(\sigma)}_\Omega u^{-i\sigma\Omega +\ell}
+ (-)^\ell
\beta^{(\sigma)}_\Omega u^{i\sigma\Omega -\ell}
\right)\nn\\
&&=\Half\int_0^\infty\frac{du}{u}
\exp\left\{\frac{i}{2}\mt\oxi\left(u-\frac{1}{u}\right)
\right\}\nn\\
&&\quad\times
\prod_n
\sum_{L_n=0}^\infty \frac{1}{L_n!}
\left[\left(\frac{u}{D_n e^{\oeta_n-\oeta}}
+\frac{D_n e^{\oeta_n-\oeta}}{u}\right)C_n y_n\right]^{L_n}
\sum_{p_n=0}^\infty
\frac{(-)^{p_n}}{p_n!}
C_n^{2p_n}\nn\\
&&\quad\times\left\{
\alpha^{(\sigma)}_\Omega u^{-i\sigma\Omega}\exp\left(
n\left(L_n+2p_n\right)\frac{i\pi\oxi}{\mt}u\right)
+
\beta^{(\sigma)}_\Omega u^{i\sigma\Omega}\exp\left(
-n\left(L_n+2p_n\right)\frac{i\pi\oxi}{\mt}\frac{1}{u}\right)
\right\}\nn\\
&&=\Half\int_0^\infty\frac{du}{u}
\exp\left\{\frac{i}{2}\mt\oxi\left(u-\frac{1}{u}\right)
\right\}\nn\\
&&\quad\times
\Biggl\{
\alpha^{(\sigma)}_\Omega u^{-i\sigma\Omega}
\prod_n \exp\left[
\left(\frac{u}{D_n e^{\oeta_n-\oeta}}
+\frac{D_n e^{\oeta_n-\oeta}}{u}\right)C_n y_n
e^{\frac{i\pi n\oxi}{\mt}u}
- C_n^2 e^{\frac{2i\pi n\oxi}{\mt}u}\right]\nn\\
&&\qquad + \beta^{(\sigma)}_\Omega u^{i\sigma\Omega}
\prod_n \exp\left[
\left(\frac{u}{D_n e^{\oeta_n-\oeta}}
+\frac{D_n e^{\oeta_n-\oeta}}{u}\right)C_n y_n
e^{-\frac{i\pi n\oxi}{\mt}\frac{1}{u}}
- C_n^2 e^{-\frac{2i\pi n\oxi}{\mt}\frac{1}{u}}\right]
\Biggr\}\ ,
\end{eqnarray}
where the summation over $m_n$ was carried out using the binomial
theorem, and we removed the regularization factor for $Q_\nu(x)$
(\ref{eq:Q}).

\newpage
\newcommand{\J}[4]{{\sl #1} {\bf #2} (19#4) #3}
\newcommand{\andJ}[3]{{\bf #1} (19#3) #2}
\newcommand{\MPL}{Mod.\ Phys.\ Lett.}
\newcommand{\NP}{Nucl.\ Phys.}
\newcommand{\PL}{Phys.\ Lett.}
\newcommand{\PR}{Phys.\ Rev.}
\newcommand{\PRL}{Phys.\ Rev.\ Lett.}
\newcommand{\AP}{Ann.\ Phys.}
\newcommand{\CMP}{Commun.\ Math.\ Phys.}
\newcommand{\PTP}{Prog.\ Theor.\ Phys.}

\end{document}